\documentclass[aps,showpacs,twocolumn,pra,a4paper,floatfix,nofootinbib]{revtex4-1}
\usepackage{amsmath}
\usepackage{amssymb}
\usepackage{amsfonts}
\usepackage{fancybox}
\usepackage{eepic}
\usepackage{times}
\usepackage{latexsym}
\usepackage{pifont}
\usepackage{graphicx}
\usepackage{amstext}

\def\acetylene{C$_2$H$_2$}
\def\jpbo{J. Phys. B: At. Mol. Opt. Phys.}   

\def\pra{Phys. Rev. A}

\def\prl{Phys. Rev. Lett.}

\def\jcp{J. Chem. Phys.}        
\def\prx{Phys. Rev. X}
\newcommand{\bv}[1]{\mbox{\boldmath $#1$}}
\newcommand{\intensity}[2]{$#1\times10^{#2}\mbox{ W/cm}^{2}$}
\def\md{\mathrm{d}}
\def\acetylene{C$_2$H$_2$}
\newcommand{\htwop}{H$_2^+$}

\begin{document}
\title{\textbf{Probing the role of excited states in ionization of acetylene}}
\author{Daniel Dundas, Peter Mulholland, Abigail Wardlow and Alejandro de la Calle}
\affiliation{School of Mathematics and Physics, Queen's University Belfast, University
Road, Belfast, BT7 1NN, N. Ireland, UK.}
\date{\today}

\begin{abstract}
Ionization of acetylene by linearly-polarized, vacuum ultraviolet (VUV) laser pulses 
is modelled using time-dependent density functional theory. Several laser wavelengths are considered including 
one that produces direct ionization to the first excited cationic state while another excites the molecules
to a Rydberg series incorporating an autoionizing state. We show that for the wavelengths and intensities 
considered, ionization is greatest whenever the molecule is aligned along the laser polarization direction. 
By considering high harmonic generation we show that populating excited states can lead to a large enhancement 
in the harmonic yield. Lastly, angularly-resolved photoelectron spectra are calculated which show how the energy 
profile of the emitted electrons significantly changes in the presence of these excited states.
\end{abstract}

\maketitle

\section{Introduction}

Attoscience is a field that has seen spectacular progress over the last several years with the continuing 
development of attosecond laser sources with which to probe dynamical processes in molecules, nanoscale
devices and solids~\cite{krauz:2009}. Since electrons move on the attosecond timescale, 
one of the main goals of attosecond chemistry is steering reactions through correlated electronic 
processes~\cite{lepine:2014}. Realizing this goal is key to the development of future ultrafast 
technologies: examples include the design of electronic devices~\cite{joachim:2000}, probes and 
sensors~\cite{willard:2003}, biological repair and signalling processes and development of 
optically-driven ultrafast electronics~\cite{krauz:2014}. Studying attosecond processes in molecules involves 
initiating a reaction by driving electrons far from equilibrium using a pump laser: one of the most widely 
used techniques is to ionize the molecule. The subsequent evolution of the electronic dynamics is followed by 
probing the system at later times using methods such as high-harmonic spectroscopy and photoelectron 
spectroscopy~\cite{lepine:2014,calegari:2014,trabattoni:2015,marangos:2016}. Unravelling these dynamical processes presents a huge 
challenge due to the number of potential pathways that exist as well as the inherent difficulties of relating 
the initial and intermediate states to the photo-products produced. For this reason studying small molecules has 
attracted a lot of effort since they represent a compromise between having sufficient complexity and limiting the 
number of photo-products when compared to much larger molecules.

Acetylene (\acetylene) is a small, linear polyatomic molecule which is isolelectronic to N$_2$. It has the ground 
configuration $(1\sigma_g)^2(1\sigma_u)^2(2\sigma_g)^2(2\sigma_u)^2(3\sigma_g)^2(1\pi_u)^4$. In addition, the
next lowest unoccupied orbitals  are $(1\pi_g)^0(3\sigma_u)^0(4\sigma_g)^0$. Ionization of acetylene has been studied 
both theoretically and experimentally by many groups. This is in part due to its rich electronic structure and in part 
due to its importance in many areas of chemistry. Initial studies of acetylene considered ionization by 
electron~\cite{collin:1967,king:2007} and 
photon~\cite{collin:1967,langhoff:1981,machado:1982,holland:1983,yasuike:2000,fronzoni:2003} impact. More recently, several 
studies have considered the response to strong fields. This has included single~\cite{argenti:2012,ji:2015} and 
double~\cite{gaire:2014,gong:2014} ionization, dissociative ionization of highly excited states~\cite{zamith:2003}, 
high harmonic generation~\cite{vozzi:2010,torres:2010,vozzi:2012,negro:2014} and more recently initiating isomerization 
of acetylene to vinylidene~\cite{ibrahim:2014}. 

The theoretical and computational study of molecules interacting with intense, ultrashort pulses is highly 
demanding. Several techniques have been developed to model laser-molecule interactions. These include simple yet 
powerful techniques like the strong field approximation (SFA)~\cite{lewenstein:1994,spanner:2009,smirnova1:2009} and 
quantitative rescattering (QRS) theory~\cite{le:2009,lin:2010} for modelling high-harmonic generation (HHG) in complex 
molecules. However, these approaches suffer from the drawback that only a small number of channels are generally 
included in a calculation. This can become an issue, depending on the molecule under investigation, if other 
channels become important during the interaction with the laser. In order to overcome these shortcomings,
a range of techniques, based on ab initio methods, have been developed. Pioneering ab initio studies using
density functional theory (DFT)~\cite{fronzoni:2003,remacle:2006,argenti:2012,calegari:2014} and configuration interaction (CI) 
methods~\cite{lunnermann:2008} have been used to study electron migration in molecules following sudden 
removal of an electron and have observed charge oscillations along the full length of the molecule 
with a period of several femtoseconds. The drawback of these approaches is that the initial pumping of the
molecule by a laser pulse is not described. In light of this several approaches that describe the explicit interaction 
with the laser field have been developed~\cite{thachuk:1996,mitric:2009,richter:2011}.

One ab initio method that is widely used to treat electronic dynamics is time-dependent density functional 
theory (TDDFT)~\cite{runge:1984}. The TDDFT method has been extensively applied to the study of laser-molecule 
and laser-cluster interactions~\cite{kunert:2003,calvayrac:2000,marques:2003}. Since we generally want to include
fragmentation processes in molecules we must couple the treatment of the electronic subsystem to another approach 
for treating the ionic dynamics. One popular approach is the Ehrenfest method whereby the quantum treatment 
of the electrons is coupled to a classical treatment of the ions, resulting in a method known as non-adiabatic 
quantum molecular dynamics (NAQMD)~\cite{kunert:2003}. 

In this paper we study ionization processes in acetylene using this mixed quantum-classical method. The paper is 
laid out as follows. In section~\ref{sec:theory} we describe our method and show how it is applied to the current 
study. In particular we describe how we calculate observables such as photoelectron spectra (PES). In 
section~\ref{sec:results} we apply our method to study ionization of acetylene by VUV laser pulses. After obtaining the 
equilibrium ground state we firstly consider how the wavelength and orientation of the laser pulse alters ionization. Then 
we show that resonant excitation of particular excited states can dramatically enhance HHG. Lastly, we calculate 
angularly-resolved PES for acetylene at two distinct wavelengths and show how excited states significantly alter the emitted
electrons. Finally, some conclusions and directions for future work are given in section~\ref{sec:conclusions}.

\section{Theoretical Approach}
\label{sec:theory}
In our calculations we implement the NAQMD method in a code called EDAMAME (Ehrenfest DynAMics on Adaptive MEshes) which is 
described in more detail in previous papers~\cite{dundas:2012,wardlow:2016}. We now briefly describe how it is applied to the 
calculations considered here. In sections~\ref{sec:theory_elec} and~\ref{sec:theory_ions} we set out the details of the method
for treating the electrons and ions respectively. Section~\ref{sec:theory_numerical} gives information on the numerical 
implementation of the approach, highlighting many of the parameters used for the calculations presented. 
Section~\ref{sec:theory_observables} describes how we calculate observables such as the amount of ionization, the HHG
spectra and angularly-resolved photoelectron spectra.

\subsection{TDDFT description of the electronic dynamics}
\label{sec:theory_elec}
We consider a molecule consisting of $N_e$ electrons and $N_n$ classical ions where 
$M_k$, $Z_k$ and $\bv{R}_k$ denote respectively the mass, charge and position of ion $k$. Additionally 
$\bv{R} = \{\bv{R}_1, \dots, \bv{R}_{N_n}\}$. Neglecting spin effects, the time-dependent electron density for the 
electrons can be expressed in terms of $N = N_e / 2$ time-dependent Kohn-Sham orbitals, $\psi_{j} (\bv{r},t)$, each 
having an initial occupancy of two, as
\begin{equation}
n(\bv{r},t) =  2\sum_{j=1}^{N} \left | \psi_{j } (\bv{r},t)  \right |^2.
\end{equation}
These Kohn-Sham orbitals satisfy the time-dependent Kohn-Sham equations (TDKS)
\begin{align}
i\frac{\partial }{\partial t}\psi_{j} (\bv{r},t) & = 
\Biggr [ 
-\frac{1}{2} \nabla^2 + V_{\mbox{\scriptsize H}}(\bv{r}, t)+
V_{\mbox{\scriptsize ext}}(\bv{r}, \bv{R}, t)+ V_{\mbox{\scriptsize xc}}(\bv{r},  t) \Biggr] \psi_{j} (\bv{r},t)
\nonumber\\
& 
 = 
H_{\mbox{\scriptsize ks}}\psi_{j} (\bv{r},t) \hspace*{2cm}j = 1,\dots, N
.
\label{eq:tdks}
\end{align}
In Eq.~(\ref{eq:tdks}), $V_{\mbox{\scriptsize H}}(\bv{r}, t)$ is the Hartree potential, 
$V_{\mbox{\scriptsize ext}}(\bv{r}, \bv{R}, t)$ is the external potential, and 
$V_{\mbox{\scriptsize xc}}(\bv{r},  t)$ is the exchange-correlation  potential. Both the Hartree and
exchange-correlation potentials are time-dependent due to their functional dependence on the
time-dependent density.
The external potential accounts for both electron-ion interactions and the interaction of the 
laser field with the electrons and can be written as
\begin{equation}
V_{\mbox{\scriptsize ext}}(\bv{r}, \bv{R}, t) =
V_{\mbox{\scriptsize ions}}(\bv{r}, \bv{R}, t) +
U_{\mbox{\scriptsize elec}}(\bv{r}, t).
\end{equation}
For the laser pulses considered in this paper, the innermost electrons can be considered frozen and so we
need only consider the response of the valence electrons. In that case, we describe the electron-ion interactions, 
$V_{\mbox{\scriptsize ions}}(\bv{r}, \bv{R}, t)$, with Troullier-Martins 
pseudopotentials~\cite{Troullier:1991} in the Kleinman-Bylander form~\cite{Kleinman:1982}. 
The pseudopotentials were generated using the Atomic Pseudopotential 
Engine (APE)~\cite{Oliveira:2008}. 

All calculations are carried out using the dipole approximation. Within EDAMAME, both length and velocity gauge 
descriptions of the electron-laser interaction, $U_{\mbox{\scriptsize elec}}(\bv{r}, t)$, can be
used. For the length gauge we have  
\begin{equation}
U_{\mbox{\scriptsize elec}}(\bv{r},t) = U_L (\bv{r},t) = \bv{r} \cdot \bv{E}(t),
\end{equation}
while for the velocity gauge
\begin{equation}
U_{\mbox{\scriptsize elec}}(\bv{r},t) = U_V (\bv{r},t) = -i\bv{A}(t) \cdot \bv{\nabla}.
\end{equation}
In these equations the vector potential, $\bv{A}(t)$, is related to the electric field, $\bv{E}(t)$, through the 
relation ${\bf E}(t) = -\frac{\partial}{\partial t}{\bf A}(t)$. Therefore, for the linearly polarised laser pulses 
used in this work we consider a $\sin^2$ pulse envelope and write
\begin{equation}
\label{eq:vector_potential}
\bv{A}(t) = A_0 \sin^2\left(\frac{\pi t}{T}\right) \cos(\omega_L t)\hat{{\bv{e}}}.
\end{equation}
Here, $A_0$ is the peak value of the vector potential, $\omega_L$ is the laser frequency, $T$ is the pulse 
duration and $\hat{\bv{e}}$ is the unit vector in the polarization direction of the laser field. 
From this, the electric vector is
\begin{equation}
\bv{E}(t)= E(t)\bv{\hat{e}},
\end{equation}
where 
\begin{equation}
E(t) = E_0 f(t) \sin(\omega_L t) - 
\frac{E_0}{\omega_L}\frac{\partial f}{\partial t} \cos (\omega_L t),
\end{equation}
and where $E_0$ is the peak electric field strength. We find good agreement between the solutions
using both gauges. However, using the length gauge is computationally less expensive than the velocity 
gauge and so in most calculations we will use the length gauge. We will use the velocity gauge for 
calculating the photoelectron spectra using the t-SURFF method (our implementation of t-SURFF in EDAMAME is
described later).

Within TDDFT all electron correlation effects are included through the exchange-correlation potential, 
$V_{\mbox{\scriptsize xc}}(\bv{r},  t)$, which is derived from an exchange-correlation action functional. 
This functional is unknown and so must be approximated. In addition it contains information on the whole 
history of past densities. In order to avoid these memory effects the adiabatic approximation is generally 
used so that ground-state approximations to the exchange-correlation functional can be employed~\cite{burke:2005}. 
In our calculations we use the local density approximation (LDA) incorporating the Perdew-Wang 
parameterization of the correlation functional~\cite{Perdew:1992}. The LDA describes the 
electrons as a homogeneous electron gas. Although this functional is widely used, it suffers from 
self-interaction errors which means that it does not have the correct long-range behaviour. One major 
consequence is that electrons are too loosely bound and excited states are not accurately 
described~\cite{casida:1998}. Therefore, we supplement this functional with the average density 
self-interaction correction (ADSIC)~\cite{legrand:2002} which reinstates the correct long-range 
behaviour in an orbital-independent fashion. This implementation will be referred to as LDA-PW92-ADSIC.
The description of autoionizing resonances arising from double excitations is problematic in TDDFT. However, 
in the present work resonance phenomena primarily arise from single excitations which are described
well using TDDFT~\cite{krueger:2009}. In particular, previous TDDFT studies of photoionization of acetylene 
have shown a good level of accuracy~\cite{fronzoni:2003}.

\subsection{Classical description of the ionic dynamics}
\label{sec:theory_ions}
The ionic dynamics are treated classically using Newton's equations of motion. For ion $k$ we
have
\begin{align}
M_k \ddot{\bv{R}}_k = &- 
\int n (\bv{r},t) \frac{\partial H_{\mbox{\scriptsize ks}}}{\partial \bv{R}_k} 
\md\bv{r} \nonumber \\
& -\frac{\partial}{\partial \bv{R}_k} \Bigr(V_{nn} (\bv{R}) + Z_k \bv{R}_k\cdot \bv{E}(t)\Bigr),
\label{eq:neom}
\end{align}
where $V_{nn} (\bv{R})$ is the Coulomb repulsion between the ions and $Z_k \bv{R}_k\cdot \bv{E}(t)$ 
denotes the interaction between ion $k$ and the laser field. A major benefit of LDA-PW92-ADSIC is that 
the exchange-correlation potential is obtained as the functional derivative of an exchange-correlation 
functional, meaning that the forces acting on ions can be derived from the Hellmann-Feynman theorem.

\subsection{Numerical implementation of the method}
\label{sec:theory_numerical}
We solve the Kohn-Sham equations numerically using finite difference techniques on a three dimensional 
Cartesian grid~\cite{dundas:2012}. All derivative operators are approximated using 9-point finite difference 
rules and the resulting grid is parallelized in 3D. Converged results were obtained using grid spacings of 
$0.4\, a_0$ for all coordinates. When using linearly-polarized laser pulses we have to treat ionizing and 
recolliding wavepackets that mainly travel along the laser polarization direction. We align the pulse
along $z$ and use a larger grid extent in this direction.

Both the TDKS equations, Eq.~(\ref{eq:tdks}), and ionic equations of motion, Eq.~(\ref{eq:neom}),  
need to be propagated in time. For the TDKS equations, this is achieved  using an 18th-order unitary 
Arnoldi propagator~\cite{dundas:2004, Arnoldi:1951, Smyth:1998}. For the ionic equations of motion, 
we propagate using the velocity-Verlet method. Converged results are obtained for a timestep of $0.2$ a.u.
for both sets of equations.

During the interaction of the molecule with the laser pulse, ionizing wavepackets can travel to the 
edge of the spatial grid, reflect and travel back towards the molecular centre. These unphysical 
reflections can be avoided using a wavefunction splitting technique that removes those wavepackets 
reaching the edges of the grid~\cite{Smyth:1998}. The splitting is accomplished using a mask function, 
$M(\bv{r})$, that splits the Kohn-Sham orbital, $\psi_j(\bv{r},t)$, into two parts
\begin{align}
\psi_j(\bv{r},t) & = M (\bv{r})\psi_j(\bv{r},t) + \left \{ 1-M(\bv{r}) \right \}\psi_j(\bv{r},t)\nonumber\\
                 & = \psi^B_j(\bv{r},t) + \psi^S_j(\bv{r},t).
\label{eq:splitting}
\end{align}
In this equation, the first term is located near the molecule and is associated with non-ionized 
wavepackets. The second term is located far from the molecule and is associated with ionizing 
wavepackets: this part is discarded in our calculations. The point at which we apply this splitting 
must be be chosen carefully to ensure only ionizing wavepackets are removed. We write the mask function 
in the form 
\begin{equation}
M(\bv{r}) = M_x(x)M_y(y)M_z(z).
\label{eq:mask_cart}
\end{equation}
Such a decomposition is most appropriate when a grid having different extents
in each direction is used. If we consider the $x$ component we can write
\begin{equation}
\label{eq:mx}
M_x(x)=\left\{\begin{matrix}
1  & \left | x \right | \leq  x_m\\[0.4cm] 
1 - \alpha \left ( \left | x \right | -x_m \right )^5 & \left | x \right | > x_m
\end{matrix}\right. ,
\end{equation}
where
\begin{equation}
\alpha = \frac{1- M_f}{(x_f - x_m)^5}.
\end{equation}
Here $x_m$ is the point on the grid where the mask starts, $x_f$ is the maximum extent of the grid in 
$x$ and $M_f$ is the value that we want the mask function to take at the edges of the grid. 
Similar descriptions are used for $M_y(y)$ and $M_z(z)$. 

\subsection{Calculation of observables}
\label{sec:theory_observables}
Within TDDFT all quantities, including observables, are functionals of the 
electronic density. As in the case of the exchange-correlation functional, the exact form of these 
functionals are unknown in many cases and approximations must be made. In this paper, we present results for
ionization, high-harmonic generation and photoelectron spectra. We describe below how these observables are calculated. 
\subsubsection{Ionization}
The exact form of this functional is unknown and so most measures of 
ionization generally are obtained using geometric properties of the time-dependent Kohn-Sham
orbitals~\cite{ullrich:2000}. In this approach, bound- and continuum-states are separated into 
different regions of space through the introduction of an analysing box. In principal the continuum
states occupy the regions of space where the wavefunction splitting is applied and hence the 
reduction in the orbital occupancies provides a measure of ionization. 

\subsubsection{High harmonic generation}
For HHG, we calculate the spectral density, $S_k(\omega)$, along the direction 
$\hat{\bv{e}}_k$ from the Fourier transform 
of the dipole acceleration~\cite{burnett:1992} 
\begin{equation}
S_k(\omega) = \left | \int e^{ i \omega t } \,\hat{\bv{e}}_k \cdot 
\ddot{\bv{d}}(t)\,\md t \right |^2,
\label{eq:hhg_all}
\end{equation}
where $\ddot{\bv{d}}(t)$ is the dipole acceleration given by 
\begin{equation}
\ddot{\bv{d}}(t) = - \int n(\bv{r},t) \bv{\nabla} H_{\mbox{\scriptsize ks}} \md\bv{r}.
\end{equation}
Additional information can be gained by calculating the response for each Kohn-Sham
orbital~\cite{chu:2016}. In that case we calculate the dipole acceleration for each state as
\begin{equation}
\ddot{\bv{d}}_j(t) = - \int n_j(\bv{r},t) \bv{\nabla} H_{\mbox{\scriptsize ks}} \md\bv{r} =
- 2\int \left|\psi_j(\bv{r}, t)\right|^2 \bv{\nabla} H_{\mbox{\scriptsize ks}} \md\bv{r}.
\label{eq:hhg_states}
\end{equation}
While this neglects interferences
between different orbitals in the overall harmonic signal, it does give an indication of the
the contribution of that state.

\subsubsection{Photoelectron Spectra}
Photoelectron spectra are calculated using the time-dependent surface flux (t-SURFF) method. t-SURFF 
has been developed by Tao and Scrinzi for one and two electron systems~\cite{Tao:2012,Scrinzi:2012}, and 
it has been further applied to the strong field ionization of \htwop{}~\cite{Yue:2014,Yue:2013}, and 
more complex systems~\cite{Karamatskou:2014,wopperer:2016}. In this method we think of configuration 
space as being divided into two regions, divided by a spherical surface at radius $R_S$. This is similar to
the situation described by Eq.~(\ref{eq:splitting}), with the splitting occurring at $R_S$. For a sufficiently long 
time, $T$, after the interaction with the laser pulse, we assume that the inner region only contains the
bound state wavepackets whereas the outer region only contains the ionized wavepackets. The scattering
amplitudes, $b_j(\bv{k})$, for a given Kohn Sham orbital, $\psi_j(\bv{r}, T)$, can then be obtained by
projecting onto scattering solutions, $\chi_{\bv{k}}$.
These amplitudes are expressed as
\begin{align}
b_j(\bv{k})\ e^{iT\bv{k}^2/2} & = \left<\chi_{\bv{k}}\left|\psi_j(\bv{r}, T)\right.\right> = 
 \left<\chi_{\bv{k}}\left|\theta(R_S) {\left|\psi_j(\bv{r}, T)\right.}\right.\right>\nonumber\\
& 
\equiv \int_{|\bv{r}|>R_S} d\bv{r}\ \chi_{\bv{k}}^*(\bv{r}) \psi_j(\bv{r}, T). 
\label{ch5:sec1-3:volume_integral}
\end{align}

The key to the method is to convert the volume integral on the right-hand-side of 
Eq.~(\ref{ch5:sec1-3:volume_integral}) into a 
time-dependent surface integral. In order to do that, we must know the evolution of the 
wavefunction after it has passed through the surface. The Hamiltonian which rules the dynamics in the 
outer region is the Volkov Hamiltonian, which has the form
\begin{equation}
\mathcal{H}_V(t) = \frac{1}{2}\ \Biggr[ \bv{k} - \bv{A}(t) \Biggr]^2.
\end{equation}
The solutions to the corresponding TDSE are the Volkov solutions
\begin{equation}
\chi_{\bv{k}}(\bv{r}) =  \frac{1}{\left(\sqrt{2\pi}\right)^3}\ e^{i \bv{k}\cdot\bv{r}} e^{-i \Phi(\bv{k},t)},
\end{equation}
where 
\begin{equation}
\Phi(\bv{k},t) = \frac{1}{2} \int_0^t dt'\ \Biggr[ \bv{k} - \bv{A}(t') \Biggr]^2,
\end{equation}
is the Volkov phase and $\bv{A}(t)$ is the vector potential of the electric field.

In order to transform the volume integral in Eq.~(\ref{ch5:sec1-3:volume_integral}) to the time
integral of the surface flux we write
\begin{widetext}
\begin{align} 
\left<\chi_{\bv{k}}\left|\theta(R_S) {\left|\psi_j(\bv{r}, T)\right.}\right.\right> &= \int_0^T \md t\ 
\frac{\md}{\md t} \left<\chi_{\bv{k}}\left|\theta(R_S) {\left|\psi_j(\bv{r}, t)\right.}\right.\right>\nonumber \\
& = i \int_0^T \md t \Bigr<\chi_{\bv{k}}\Bigr| \mathcal{H}_V(t) \theta(R_S) - \theta(R_S) \mathcal{H}(t) 
\Bigr|\psi_j(\bv{r}, t)\Bigr>  \nonumber \\
&= i \int_0^T dt \Biggr<\chi_{\bv{k}}\Biggr| \left[ -\frac{1}{2} \nabla^2 - i \bv{A}(t) \cdot \bv{\nabla},\,
\theta(R_S) \right] \Biggr|\psi_j(\bv{r}, t)\Biggr>,
\label{eq:surface_integral}
\end{align}
\end{widetext}

\noindent
where we have introduced the Heaviside function, $\theta(x)$. We decompose the plane wave in the Volkov solutions in terms of 
spherical harmonics in $\bv{k}$- and $\bv{r}$-space as
\begin{equation} 
e^{i \bv{k}\cdot\bv{r}} = 4\pi \sum_{lm} i^l j_l(kr)
Y_{lm}(\hat{\bv{k}})Y_{lm}^\star(\hat{\bv{r}}),
\label{eq:plane-wave}
\end{equation} 
where $Y_{lm}(\hat{\bv{x}})$ is a spherical harmonic in the angular variables of
$\hat{\bv{x}}$-space for orbital angular momentum quantum number $l$ and magnetic quantum number
$m$ and $j_l(x)$ is a spherical Bessel function of the first kind.
The integrand in Eq.~(\ref{eq:surface_integral}) is then calculated according to the
formula~\cite{zielinski:2016}
\begin{widetext}
\begin{align} 
\Biggr<\chi_{\bv{k}}\Biggr| \left[ -\frac{1}{2} \nabla^2 - i \bv{A}(t) \cdot 
\bv{\nabla},\, \theta(R_S) \right] \Biggr|\psi_j(\bv{r}, t)\Biggr>  =   
\sqrt{\frac{2}{\pi}}e^{i \Phi(\bv{k},t)}R_S^2 \sum_{lm} (-i)^l
Y_{lm}(\hat{\bv{k}})& \times\nonumber\\
\left[
\begin{array}{l}
\displaystyle -\frac{1}{2}\left(-\left.\frac{\partial j_l(kr)}{\partial r}\right|_{R_S}
\psi_j(R_S, t)
+ j_l(kR_S) \left.\frac{\partial \psi_j(R_S, t)}{\partial
r}\right|_{R_S}\right)\\[0.8cm]
\displaystyle  - iA_x j_l(kR_S) \left<Y_{lm}(\hat{\bv{r}})
\left|\sin\theta\cos\phi\left|\psi_j(R_S, t)\right.\right.\right> \\[0.4cm]
\displaystyle  - iA_y j_l(kR_S) \left<Y_{lm}(\hat{\bv{r}})
\left|\sin\theta\sin\phi\left|\psi_j(R_S, t)\right.\right.\right> \\[0.4cm]
\displaystyle  - iA_z j_l(kR_S) \left<Y_{lm}(\hat{\bv{r}})
\left|\cos\theta\left|\psi_j(R_S, t)\right.\right.\right>  
\end{array}
\right].
\label{eq:tsurff_comm}
\end{align}
\end{widetext}
To implement t-SURFF, we need to know the wavefunction and its first derivative at a series of 
points on a spherical shell at $R_S$. In light of the spherical nature of the problem and the fact that 
the initial implementation of t-SURFF was in spherical coordinates, this is the most appropriate coordinate 
system to use. However, EDAMAME is written is Cartesian coordinates and so we use tri-cubic 
interpolation~\cite{lekien:2005} to interpolate the Kohn-Sham orbitals onto a spherical shell: this is carried out 
during a simulation, output to file and post-processed afterwards. To obtain the photoelectron spectra we integrate
$\left|b_j(\bv{k})\right|^2$ over all angular variables and sum up the contribution of all orbitals. We can also 
calculate angularly-resolved photoelectron spectra. In a spherical representation the angular variables of $\bv{k}$ are 
$(\theta_k, \phi_k)$. We integrate over the $\phi_k$ angle to produce plots of the spectrum for $\theta_k$ only. We 
consider two cases: one where the integration in $\phi_k$ is carried out over the positive $x$-axis, the other where it 
is carried out over the negative $x$-axis. Each of these will give a semi-circular plot which we merge together to 
form a circular plot. This allows us to take account of asymmetries in each region. This implementation of t-SURFF 
is freely available n a library called POpSiCLE (PhOtoelectron SpeCtrum library for Laser-matter 
intEractions)~\cite{popsicle}.

\begin{table*}
\begin{tabular*}{\textwidth}{@{\extracolsep{\fill}} cccccc} \hline
\multicolumn{1}{c}{} &
\multicolumn{2}{c}{Equilibrium bond lengths (Bohr)} &
\multicolumn{3}{c}{Vertical Ionization Potentials (Hartree)}\\
\multicolumn{1}{c}{} &
\multicolumn{1}{c}{C--C bond length} &
\multicolumn{1}{c}{C--H bond length} &
\multicolumn{1}{c}{$X^2 \Pi_u$} & 
\multicolumn{1}{c}{$A ^2\Sigma_g^+$} & 
\multicolumn{1}{c}{$B ^2\Sigma^+_u$} \\\cline{2-3}\cline{4-6}
\multicolumn{1}{c}{Current}      & 2.207 & 2.045 & 0.4150 & 0.5550 & 0.6118\\
\multicolumn{1}{c}{Experimental} & 2.273 & 2.003 & 0.4191 & 0.6140 & 0.6912 \\
\hline
\end{tabular*}
\caption{Static properties of acetylene. The equilibrium C--C and C--H bond lengths and ionization 
potential calculated using LDA-PW92-ADSIC are compared with experimental values. The vertical ionization 
potentials from the $X ^1\Sigma_g^+$ ground state to the $X ^2\Pi_u$, $A ^2\Sigma_g^+$ and 
$B ^2\Sigma^+_u$ cationic states are estimated, respectively, from the HOMO, HOMO-1 and HOMO-2 orbital energies.
Experimental bond lengths are taken from Yasuike \& Yabushita~\protect\cite{yasuike:2000} while the ionization 
potentials are taken from Wells \& Lucchese~\protect\cite{wells:1999}.}
\label{tab:table1}
\end{table*}

\section{Results}
\label{sec:results}
In this section we apply our method to study ionization of acetylene with ultra-short linearly-polarized VUV laser 
pulses. In particular we will study resonance phenomena arising from transitions to excited states. The section is laid 
out as follows. In section~\ref{sec:acetylene_gs} we calculate the equilibrium ground state of acetylene using our TDDFT 
method. Starting from this initial state we investigate the role of excited states in the response of acetylene 
to intense laser pulses. Firstly, in section~\ref{sec:acetylene_ionization} we show how ionization to the various cationic 
states of acetylene changes as a function of the laser wavelength and the orientation between the molecule and the laser
pulse. Secondly, in section~\ref{sec:acetylene_hhg} we show how using a VUV pulse tuned to a particular excited state 
can give rise to a large enhancement in the HHG efficiency. Finally, in section~\ref{sec:acetylene_pes} we investigate the 
role of this excited state by analysing the photoelectrons produced after interaction with VUV pulses. In particular, two 
wavelengths are considered, one of which is tuned to a particular excited state, the other which bypasses it. In the 
calculations presented here the ions are allowed to move. The effect of allowing the ions to move, when studying HHG in 
benzene, was considered in a previous paper~\cite{wardlow:2016}. For the results presented here, in which acetylene interacts with 
IR pulses, we see the hydrogen atoms oscillating with amplitudes of around 0.1 a$_0$, whereas when the molecule interacts with
VUV pulses the hydrogen atoms oscillate with amplitudes around an order of magnitude smaller.

\begin{figure}
\centerline{\includegraphics[width=5cm,viewport=20 61 313 344]{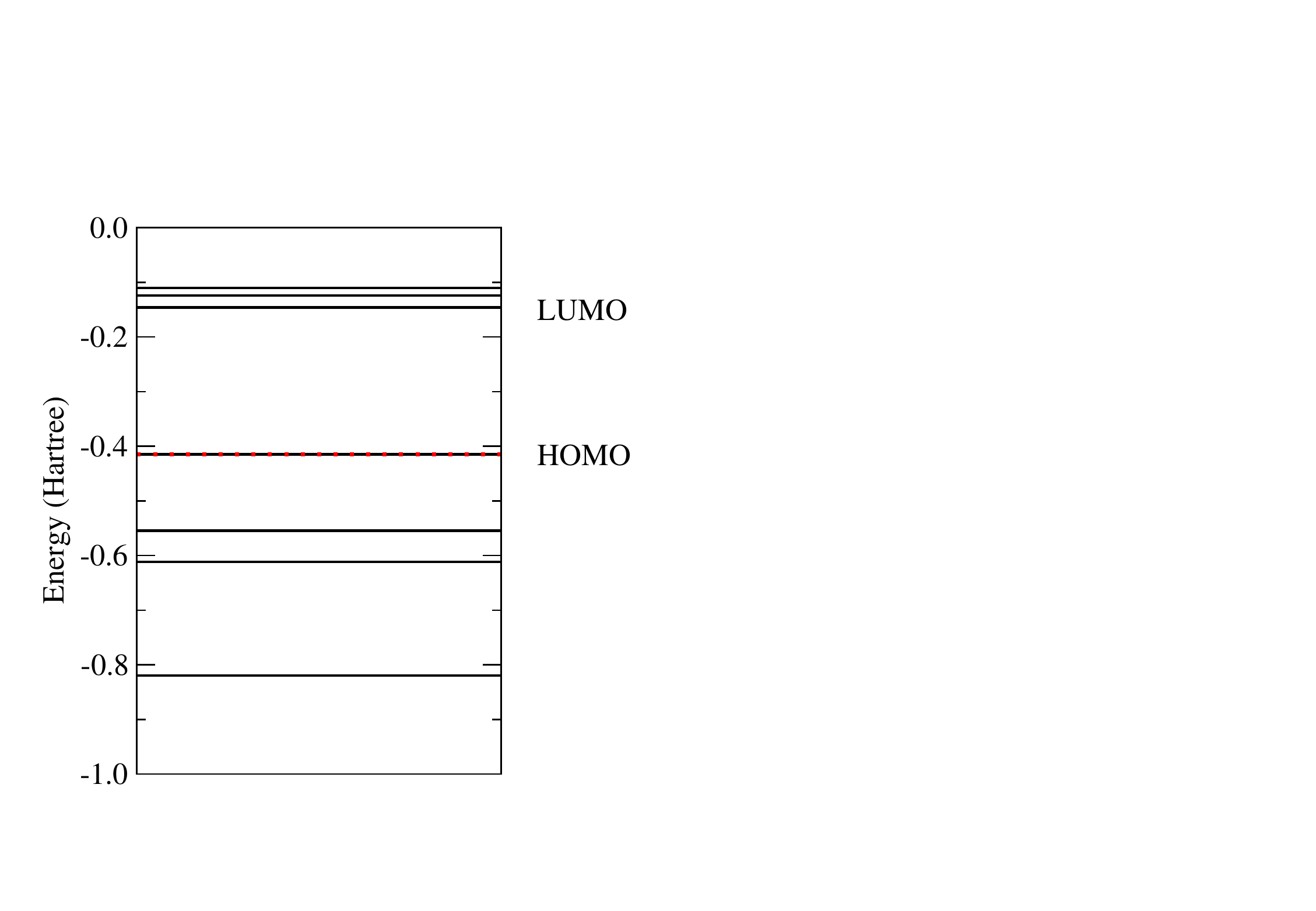}}
\caption{Occupied and unoccupied Kohn Sham orbital energies of acetylene obtained from our geometry
optimized calculation of the ground state using the ADSIC exchange-correlation functional. For
clarity the HOMO and LUMO orbitals have been marked in the figure. The red dotted line shows the
position of the negative of the experimental ionization potential of acetylene. }
\label{fig:figure1}
\end{figure}

\subsection{Equilibrium ground-state properties}
\label{sec:acetylene_gs}
The equilibrium ground state configuration of acetylene was obtained using a geometry relaxation scheme. Using the 
LDA-PW92-ADSIC exchange-correlation potential, this gave the equilibrium bond lengths and ionization potentials as shown 
in Table~\ref{tab:table1}. Using Koopman's theorem, the vertical ionization potentials are estimated from the energies of the 
Kohn-Sham orbitals. We see that these quantities are in good agreement with the experimental values. While the 
C--C bond length is well reproduced, the C--H bond length is slightly overestimated. Likewise the ionization potential to 
the ground cationic state is well reproduced using LDA-PW92-ADSIC. However, those for the next two excited cationic states 
are lower than experimental values. In addition to ionization (bound-continuum transitions), an incident laser pulse can 
also excite an electron to a bound excited state (bound-bound transitions). This corresponds to a transition from an occupied 
state to an unoccupied state. The Kohn-Sham orbital energies of the five active occupied orbitals, together with those for 
the several lowest unoccupied orbitals, are plotted in Fig.\,\ref{fig:figure1}. These energies are obtained through direct
diagonalization of the field-free Kohn-Sham Hamiltonian.

\begin{figure}
\centerline{\includegraphics[width=7cm,viewport=7 34 529 388]{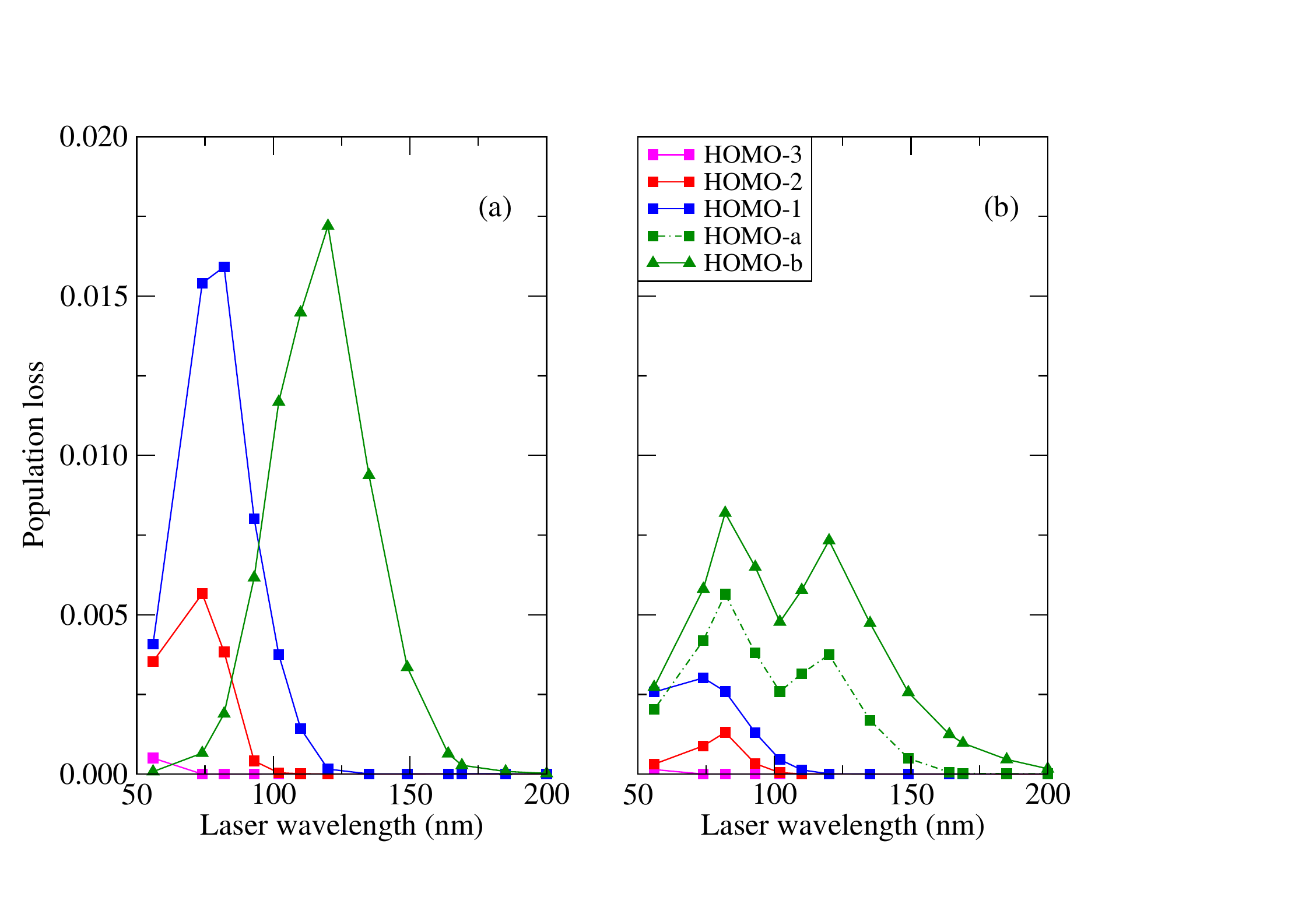}}
\caption{Population loss from the occupied Kohn-Sham states in acetylene after interaction with 8-cycle linearly polarized
VUV laser pulses having a peak intensity of $I = $ \intensity{1.0}{12}. Fourteen different wavelengths were used and two
orientations of the molecule with the laser polarization direction were considered: in plot (a) the parallel orientation is
considered while in plot (b) the perpendicular orientation is considered. Population loss from each orbital is associated with
ionization to a different cationic state. In plot (a) we only present the HOMO-a orbital since the HOMO-b orbital has the same
response. The same vertical scale is used in both plots.}
\label{fig:figure2}
\end{figure}

\subsection{Ionization to different cationic states}
\label{sec:acetylene_ionization}
The likelihood of ionization to a particular cationic state depends greatly upon the symmetry of the states involved (and thus 
the orientation of the molecule with the laser pulse) and the energy required for ionization. For example, consider ionization of
acetylene by high-intensity IR laser pulses. In that case ionization from the ground state of the neutral molecule ($X^1 \Sigma_g^+$) 
to the ground state of the cation ($X^2 \Pi_u$), i.e. removal of an electron from the HOMO orbital, is more likely when the laser 
pulse is aligned perpendicular to the molecular axis rather than when it is aligned along the molecular axis. Likewise, ionization 
to the $A ^2\Sigma_g^+$ cationic state is associated with removal of an electron from the HOMO-1 orbital and is more likely when 
the laser pulse is aligned along the molecular axis. Fig.~\ref{fig:figure2} presents the population loss from each Kohn-Sham state 
during interaction of acetylene with 8-cycle (2.67~fs) linearly polarized VUV laser pulses having a peak intensity of $I = $ 
\intensity{1.0}{12}: in total fourteen different wavelengths were considered. This population loss is calculated for two orientations 
of the molecule with the laser pulses: parallel and perpendicular. 
Population loss from each Kohn-Sham orbital is associated with vertical ionization to different cationic states. We see that more 
ionization occurs in parallel alignment. This is the opposite of the response we observe at IR wavelengths using higher laser 
intsities. In the parallel orientation, we see large increases in the amount of ionization from each Kohn-Sham orbital at the 
photon energies associated with the orbital energies. In the perpendicular case the response is different, especially for the HOMO 
orbitals. In that case we see enhancement in ionization for a range of wavelengths.

\subsection{The role of excited states in HHG}
\label{sec:acetylene_hhg}
Previous studies of photoionization in acetylene have shown that the $3\sigma_g\rightarrow 3\sigma_u$ transition is associated 
with an autoionizing state which leads to a feature in the photoionization spectrum around 0.4963~Ha in the photon 
energy~\cite{yasuike:2000,wells:1999}. Such a transition is associated with an excitation from the HOMO-1 to the LUMO+1. 
Using linearly-polarized laser pulses with the polarization aligned parallel to the molecular axis, we have recently shown that 
this excitation can lead to a significant enhancement of the plateau harmonics~\cite{mulholland:2017}. In our calculations the 
HOMO-1 to LUMO+1 energy gap is 0.4445~Ha. We consider a pump-probe scheme in which acetylene interacts with an 8-cycle (2.67~fs) 
linearly polarized VUV laser pulse having a wavelength of $\lambda=$ 102~nm (photon energy $=0.4467$ Ha) and a peak intensity of 
$I = $ \intensity{1.0}{12}. For this pulse, the bandwidth is sufficient to excite the 
$2\sigma_u\rightarrow 4\sigma_g$ transition as well. Immediately after the pump pulse, the molecule interacts with a 
5-cycle (24.2~fs) laser pulse having a wavelength of $\lambda=$ 1450~nm (photon energy $=$ 0.0314 ~Ha) and a peak 
intensity of $I = $ \intensity{1.0}{14}. For these calculations the grid extents were $|x| = |y| \leq 90.8$\,a$_0$ and  
$|z| \leq 146.8$\,a$_0$. 

\begin{figure}
\centerline{\includegraphics[width=7cm,viewport=8 33 531 393]{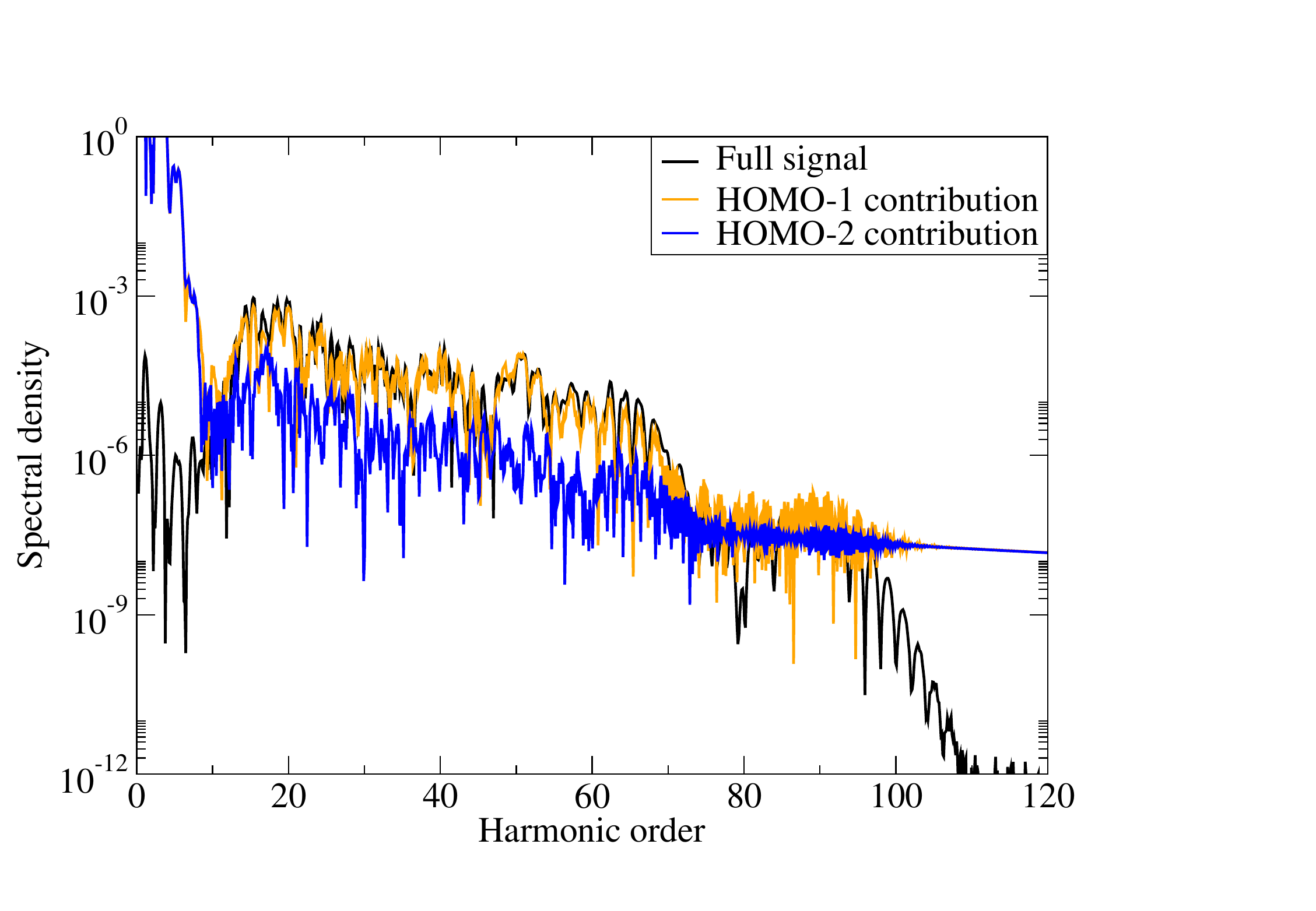}}
\caption{HHG in acetylene using a pump-probe scheme. The molecule first interacts with an 8-cycle linearly polarized
VUV laser pulse having a wavelength of $\lambda=$ 102~nm and a peak intensity of $I = $ \intensity{1.0}{12}. 
Immediately after this pulse ends the molecule interacts with 5-cycle, linearly-polarized IR laser pulse having 
a wavelength of $\lambda=$ 1450~nm and a peak intensity of $I =$ \intensity{1.0}{14}. Both the pump and probe pulses 
are aligned along the molecular axis. The full harmonic signal, calculated using Eq.~(\ref{eq:hhg_all}), is shown 
together with the contribution of the HOMO-1 and HOMO-2 orbitals, calculated using Eq.~(\ref{eq:hhg_states}).}
\label{fig:figure3}
\end{figure}

Fig.~\ref{fig:figure3} presents the harmonic response for this pump-probe scheme. We see that a double plateau structure is 
present. The outer plateau arises due to ionization from and recombination back to the HOMO-1. The inner plateau is due 
to ionization from and recombination back to the LUMO+1. However, the cut-off energy for this inner plateau is lower than 
that predicted by the classical three-step model: we attribute this to suppression of ionization from the LUMO+1 due to
transitions back to the HOMO-1~\cite{mulholland:2017}. These transitions from the LUMO+1 to the HOMO-1 give rise to a
window of enhanced harmonics (harmonics 11--21). We can show that the inner plateau is due to the
$3\sigma_g\rightarrow 3\sigma_u$ transition rather than the $2\sigma_u\rightarrow 4\sigma_g$ transition by calculating the 
contribution of each state to the harmonic response using Eq.~(\ref{eq:hhg_states}). The contributions of the HOMO-1 and 
HOMO-2 to the harmonic response are also shown in Fig.~\ref{fig:figure3}. We see that the HOMO-1 contribution dominates 
by over an order of magnitude.

\begin{figure}
\centerline{\includegraphics[width=7cm,viewport=17 34 525 391]{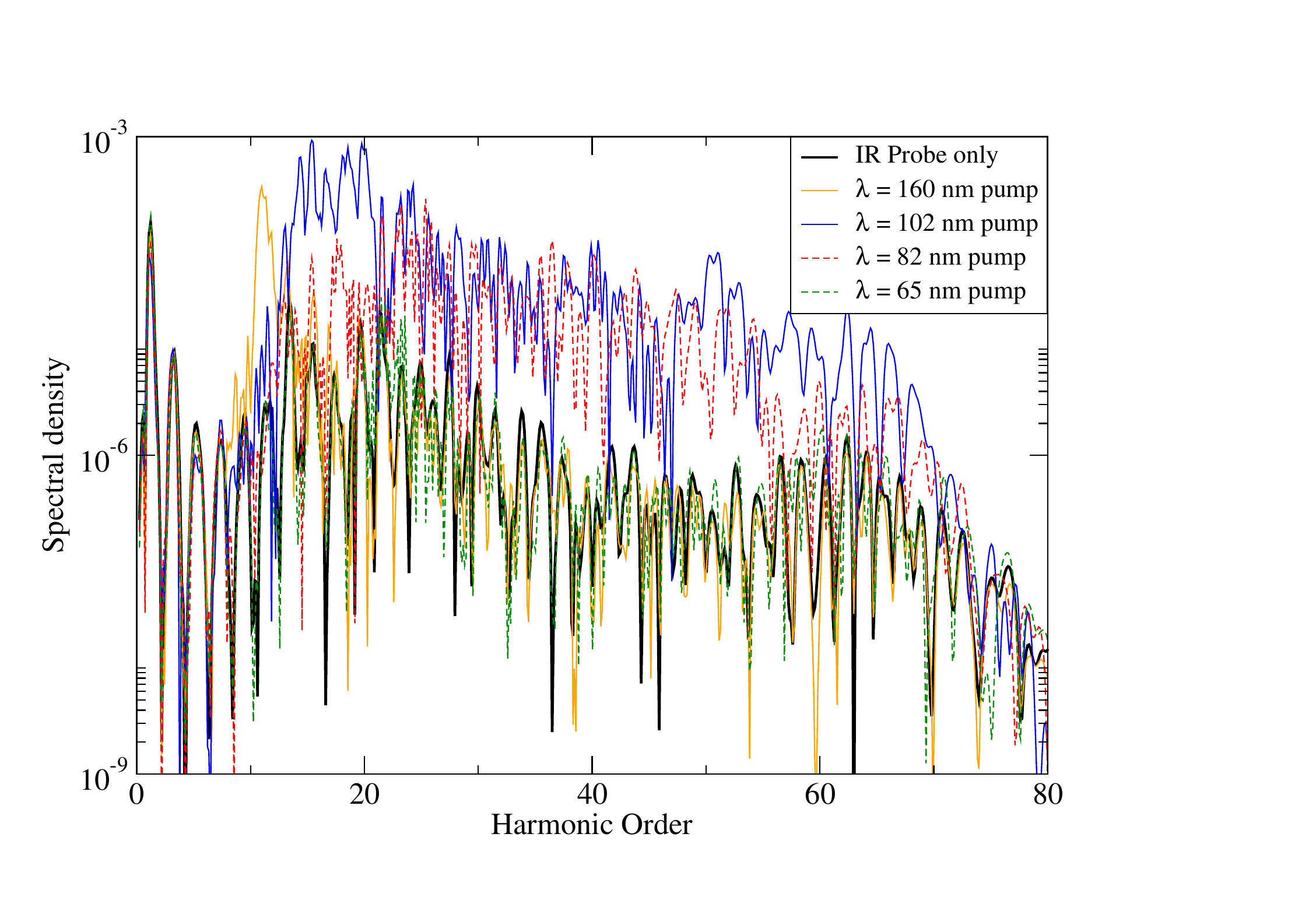}}
\caption{HHG in acetylene after interaction with a linearly-polarized VUV pump pulse followed immediately by a 
5-cycle, linearly-polarized IR laser pulse having a wavelength of $\lambda=$ 1450~nm and a peak intensity of 
$I =$ \intensity{1.0}{14}. For the pump pulse the intensity is $I =$ \intensity{1.0}{12}. Four different pump 
wavelengths were considered with the number of cycles varied to keep the total pump duration the same 
($\sim$ 2.5 fs). Both the pump and probe pulses are aligned along the molecular axis.}
\label{fig:figure4}
\end{figure}

To show that resonant excitation is the cause for the HHG enhancement, we can consider the response of 
acetylene to VUV pump pulses having different photon energies. Fig.~\ref{fig:figure4} presents results for 
four different wavelengths. In each case the laser intensity of the pump pulse is kept constant at $I=$ 
\intensity{1.0}{12} while the number of pulse cycles is changed so that the total pulse length is roughly 
the same for all four pulses ($\sim$ 2.5~fs). From this figure we see clear evidence of resonant excitation 
around $\lambda = $ 102 nm. The results for $\lambda = $ 82 nm are interesting. For this laser wavelength we are able 
to directly ionize the molecule to the $A ^2\Sigma_g^+$ excited cationic state. In that case, we would expect significant 
ionization from the $3\sigma_g$ orbital meaning that the population in the LUMO+1 (excited from the $3\sigma_g$) should 
decrease. This clearly shows 
up in the plot as a reduction in the intensity of the bound-bound harmonics (harmonics 11--21).

\begin{figure*}
\centerline{\hfill\includegraphics[width=8cm,viewport=12 34 527 391]{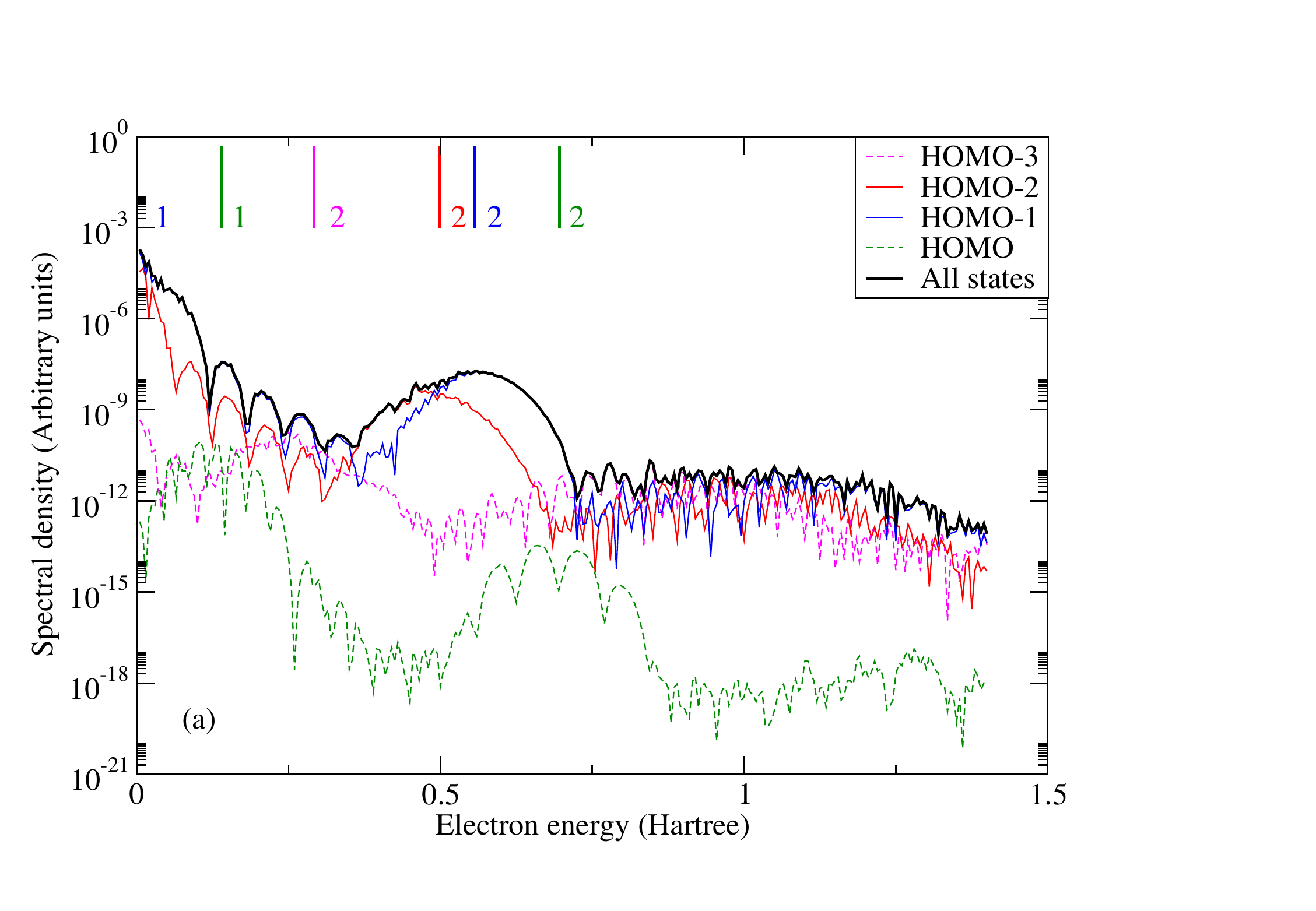}
            \hfill\includegraphics[width=8cm,viewport=12 34 527 391]{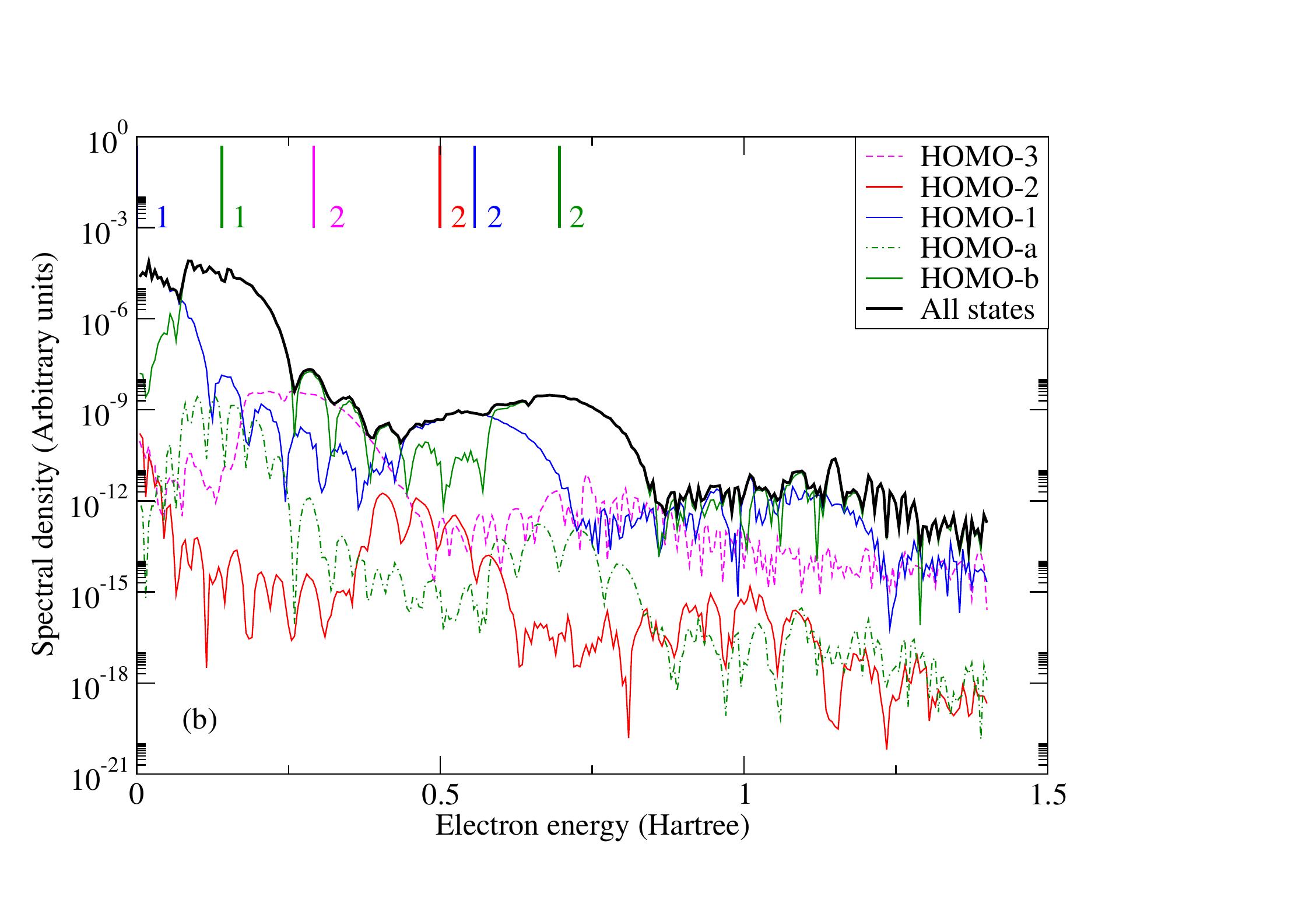}
	    \hfill}
\caption{Photoelectron spectra for acetylene after interaction with an 8-cycle linearly polarized
VUV laser pulse having a wavelength of $\lambda=$ 82~nm (photon energy $=$ 0.5557~Ha) and a peak intensity 
of $I = $ \intensity{1.0}{12}. Two orientations of the laser polarization direction with the molecular axis 
are considered. In (a) the alignment is parallel whereas in (b) the orientation is perpendicular. In the 
parallel orientation both HOMO orbitals have the same response and so we only show one. On each plot we also 
show the excess energy associated with vertical ionization due to absorption of one or two photons from each 
Kohn-Sham state.}
\label{fig:figure5}
\end{figure*}
 
\begin{figure*}
\centerline{\hfill\includegraphics[width=6cm,viewport=12 135 583 707]{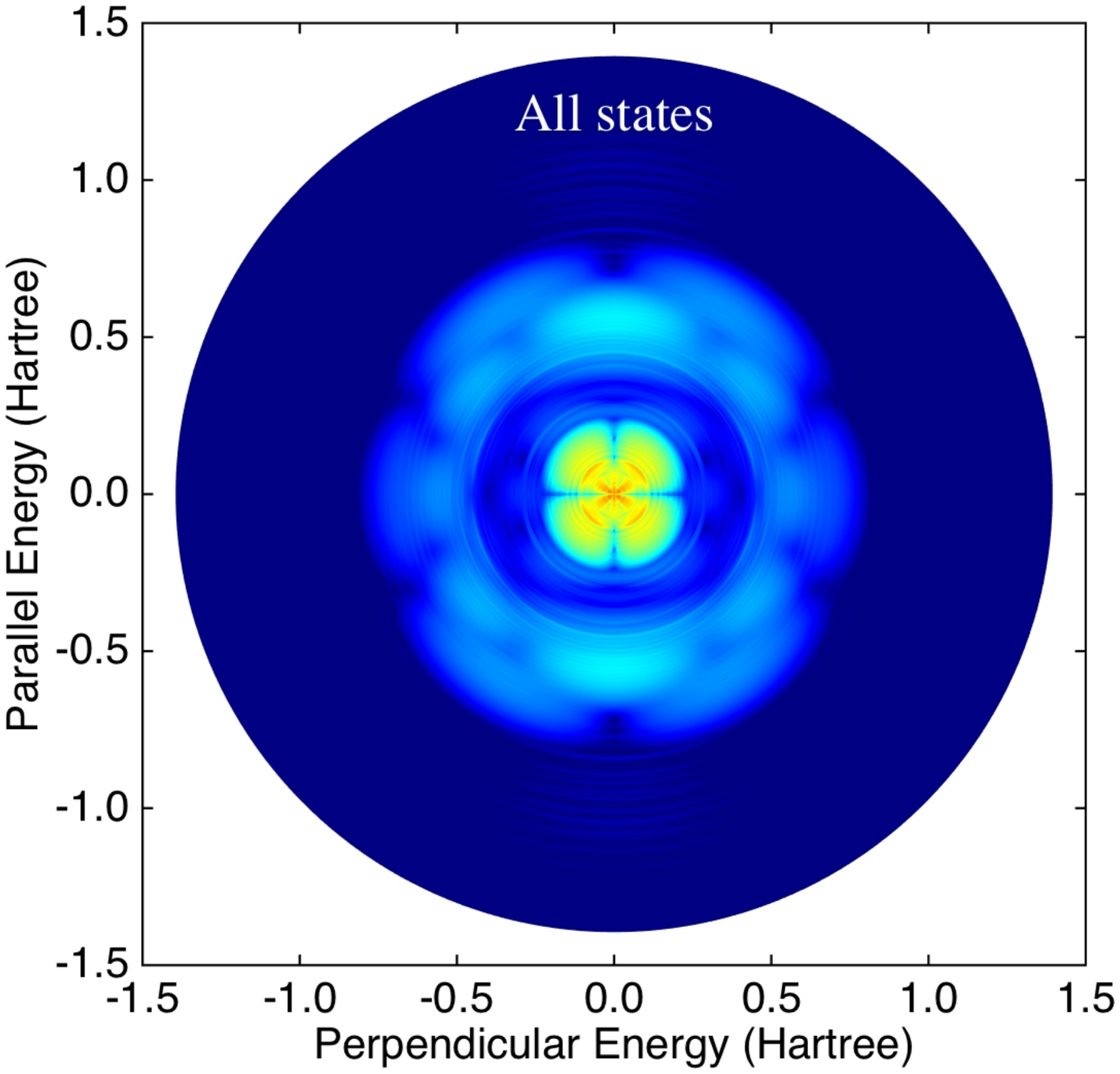}
            \hfill\includegraphics[width=6cm,viewport=12 135 583 707]{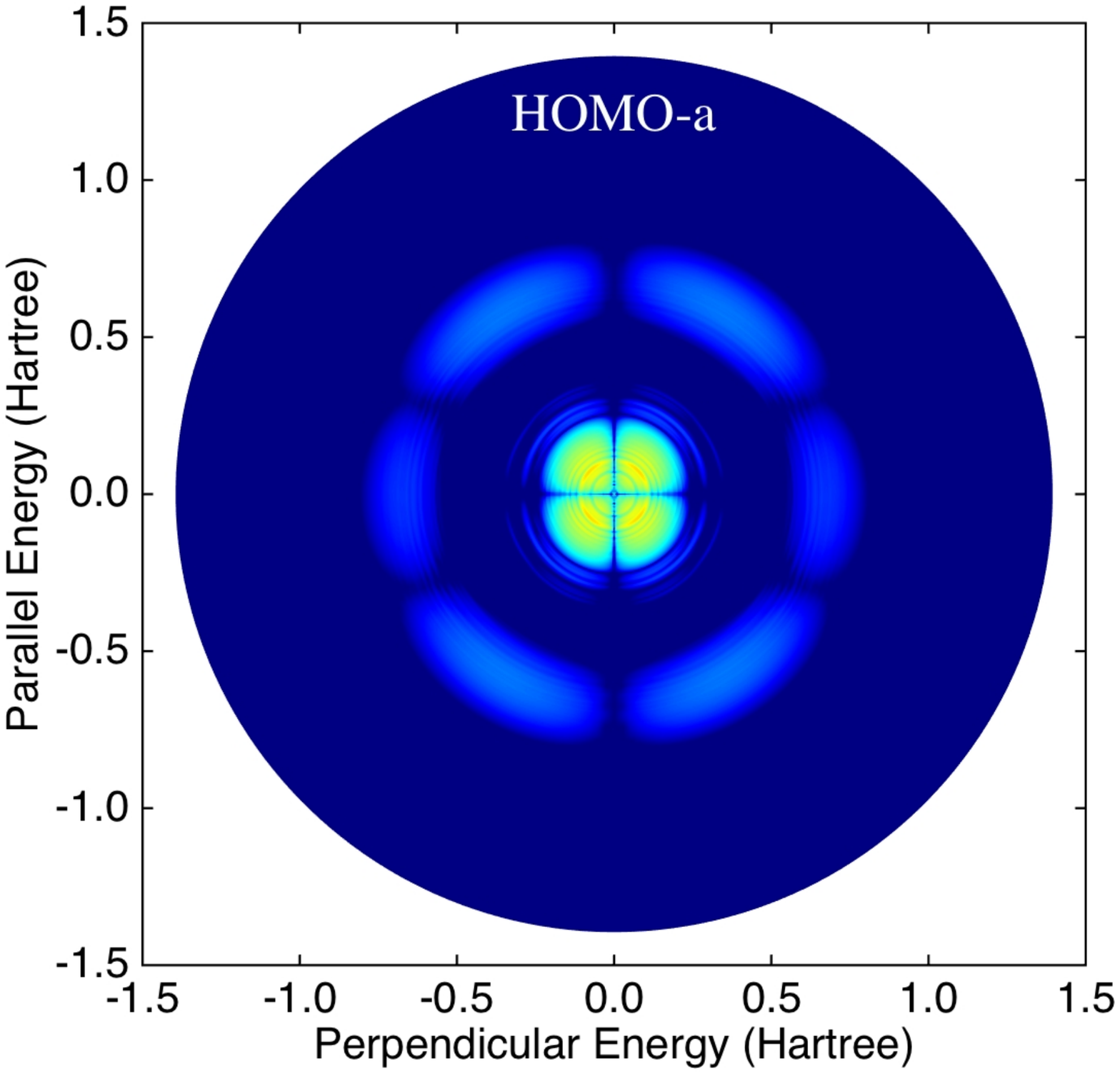}
            \hfill\includegraphics[width=6cm,viewport=12 135 583 707]{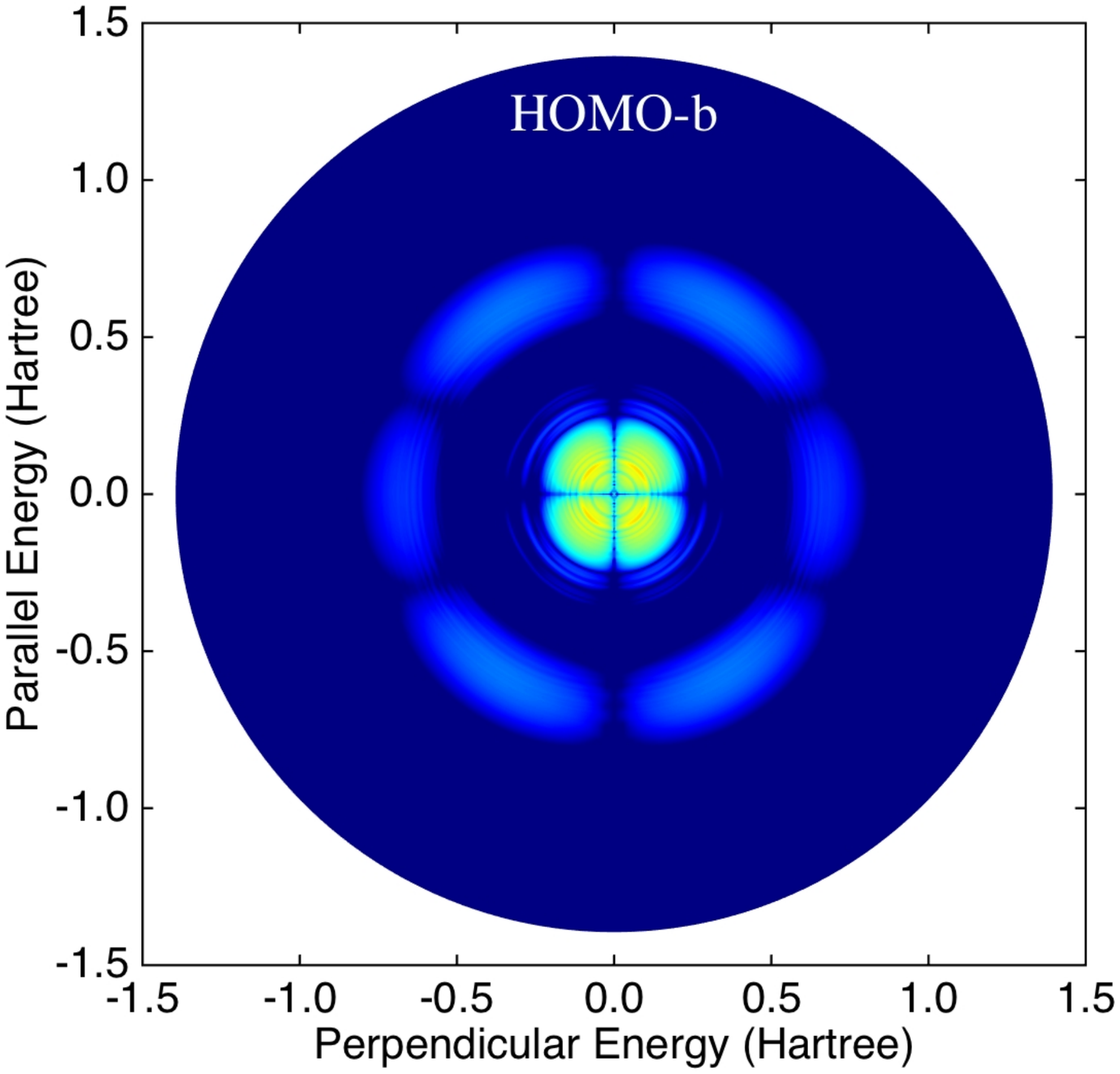}
	    \hfill}
\centerline{\hfill\includegraphics[width=6cm,viewport=12 135 583 707]{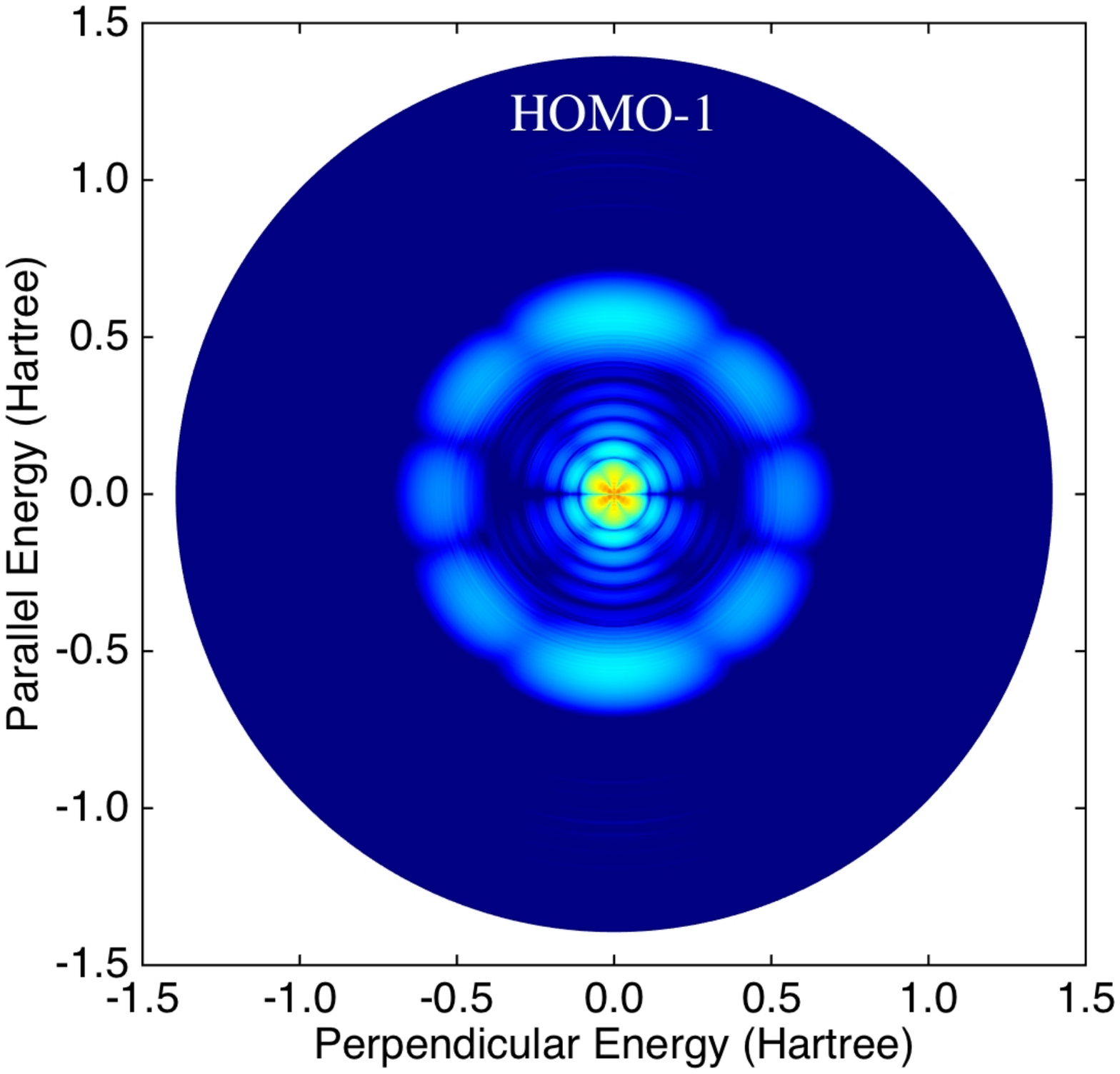}
            \hfill\includegraphics[width=6cm,viewport=12 135 583 707]{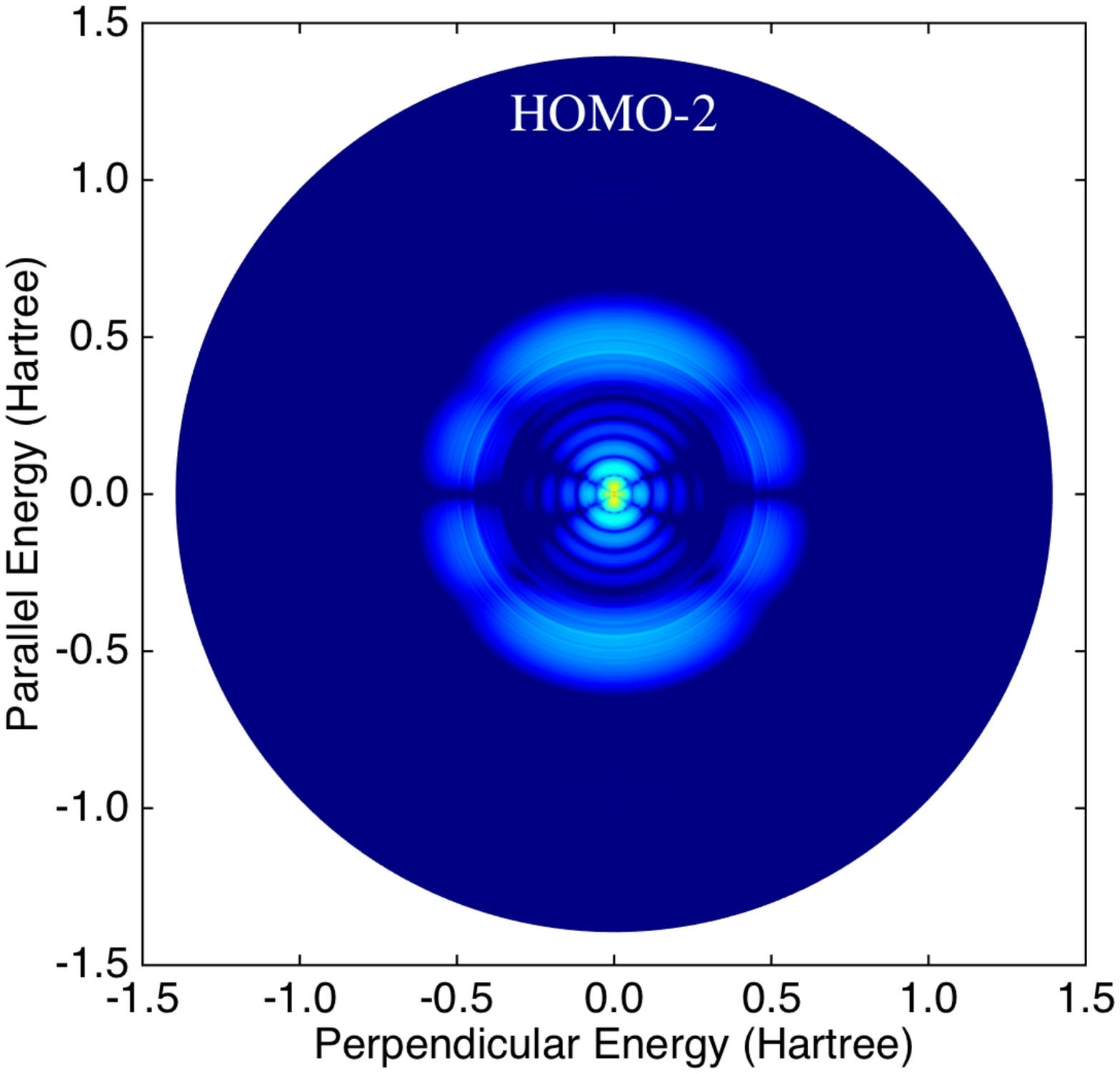}
            \hfill\includegraphics[width=6cm,viewport=12 135 583 707]{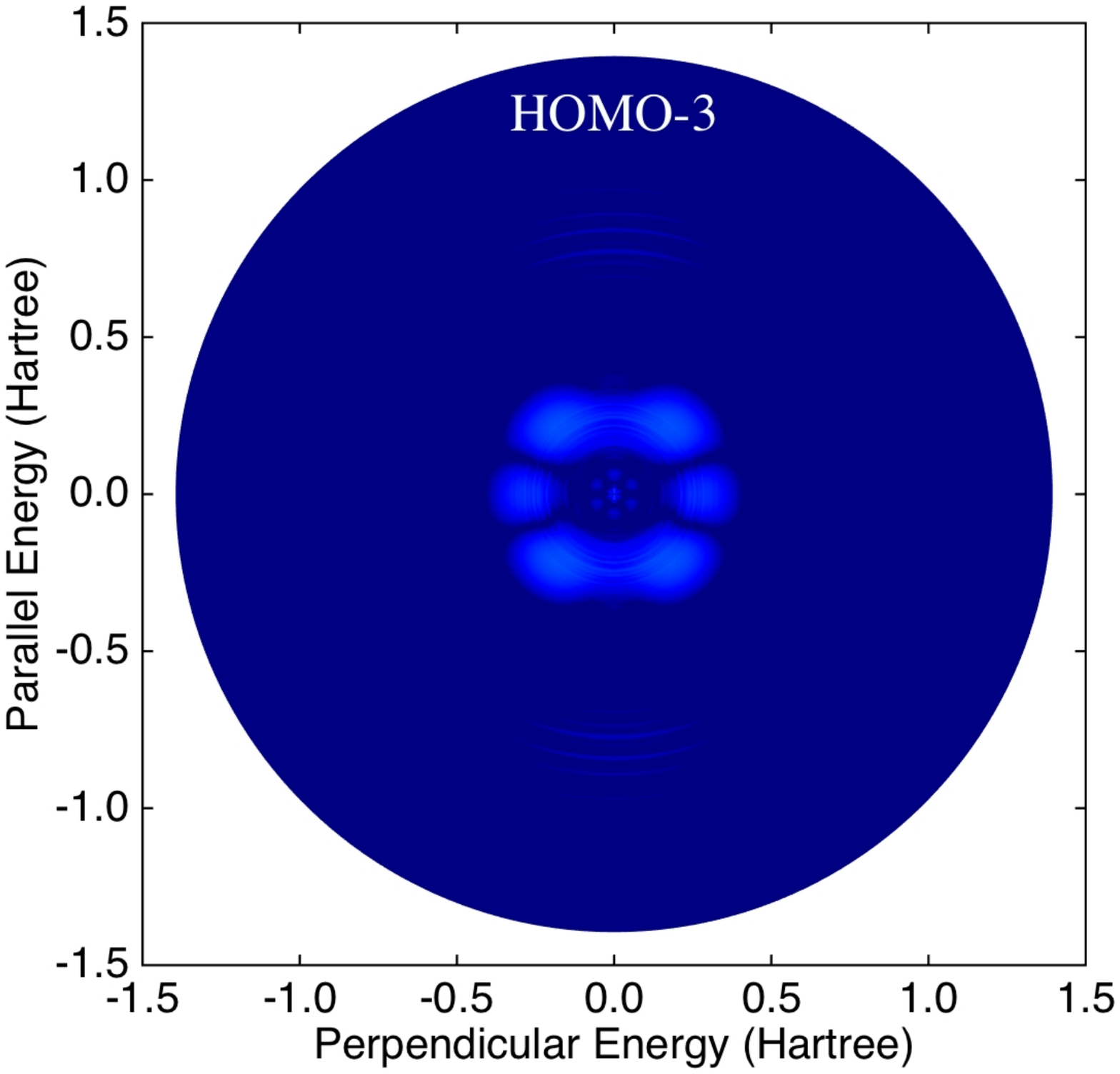}
	    \hfill}
\caption{Angularly-resolved photoelectron spectra for acetylene after interaction with an 8-cycle linearly 
polarized VUV laser pulse having a wavelength of $\lambda=$ 82~nm (photon energy $=$ 0.5557~Ha) and a peak 
intensity of $I = $ \intensity{1.0}{12}. The laser polarization direction is parallel to the molecular axis 
(along the $z$-axis). The full photoelectron spectra in 3D has been integrated in the 
azimuthal direction for both the negative $x$-axis and the positive $x$-axis. The left semi-circle of each plot 
gives the results for the negative $x$-axis while the right semi-circle of each plot gives the results 
for the positive $x$-axis. Due to symmetry both quadrants give the same response.}
\label{fig:figure6}
\end{figure*}
\begin{figure*}
\centerline{\hfill\includegraphics[width=6cm,viewport=12 135 583 707]{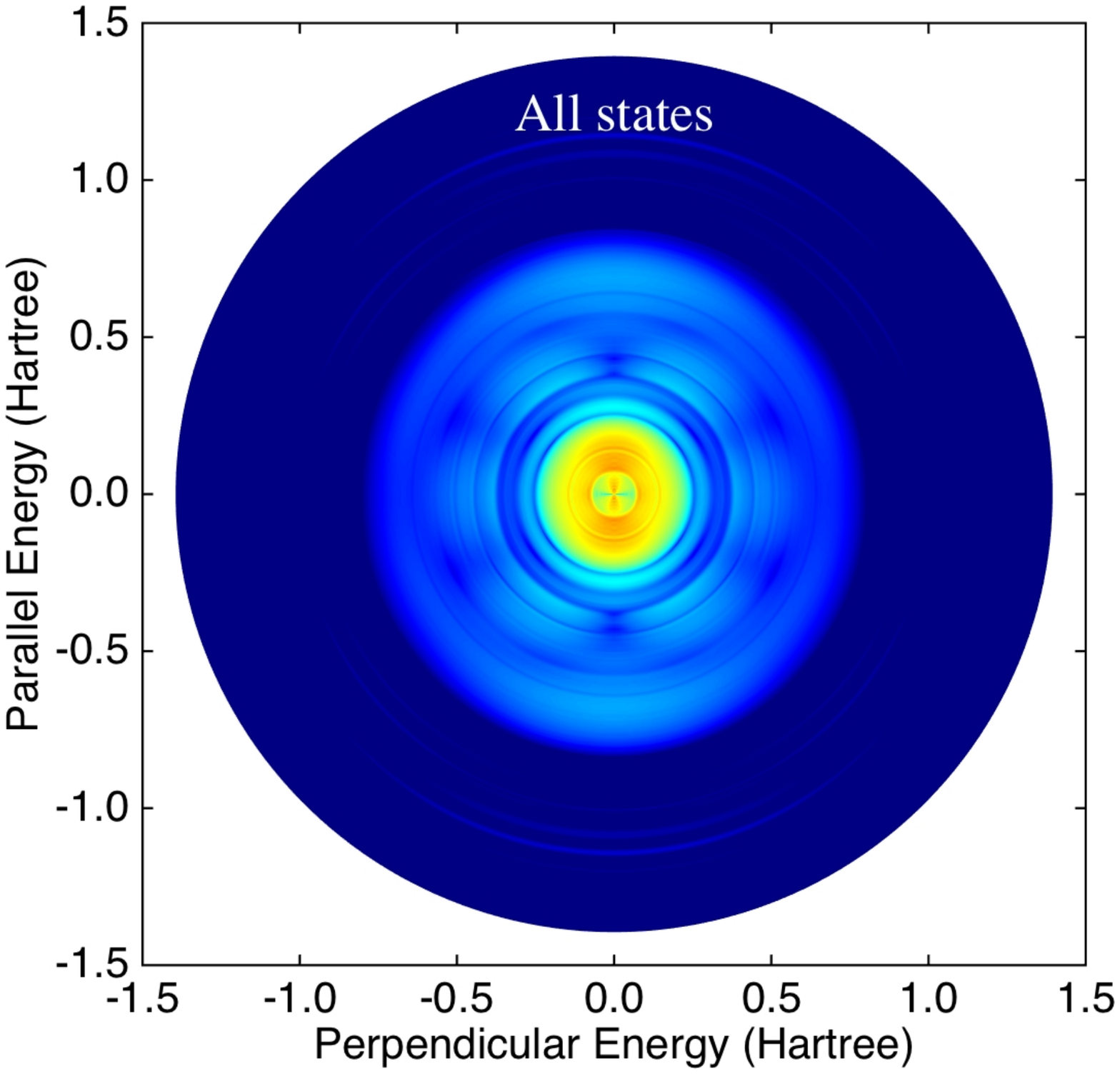}
            \hfill\includegraphics[width=6cm,viewport=12 135 583 707]{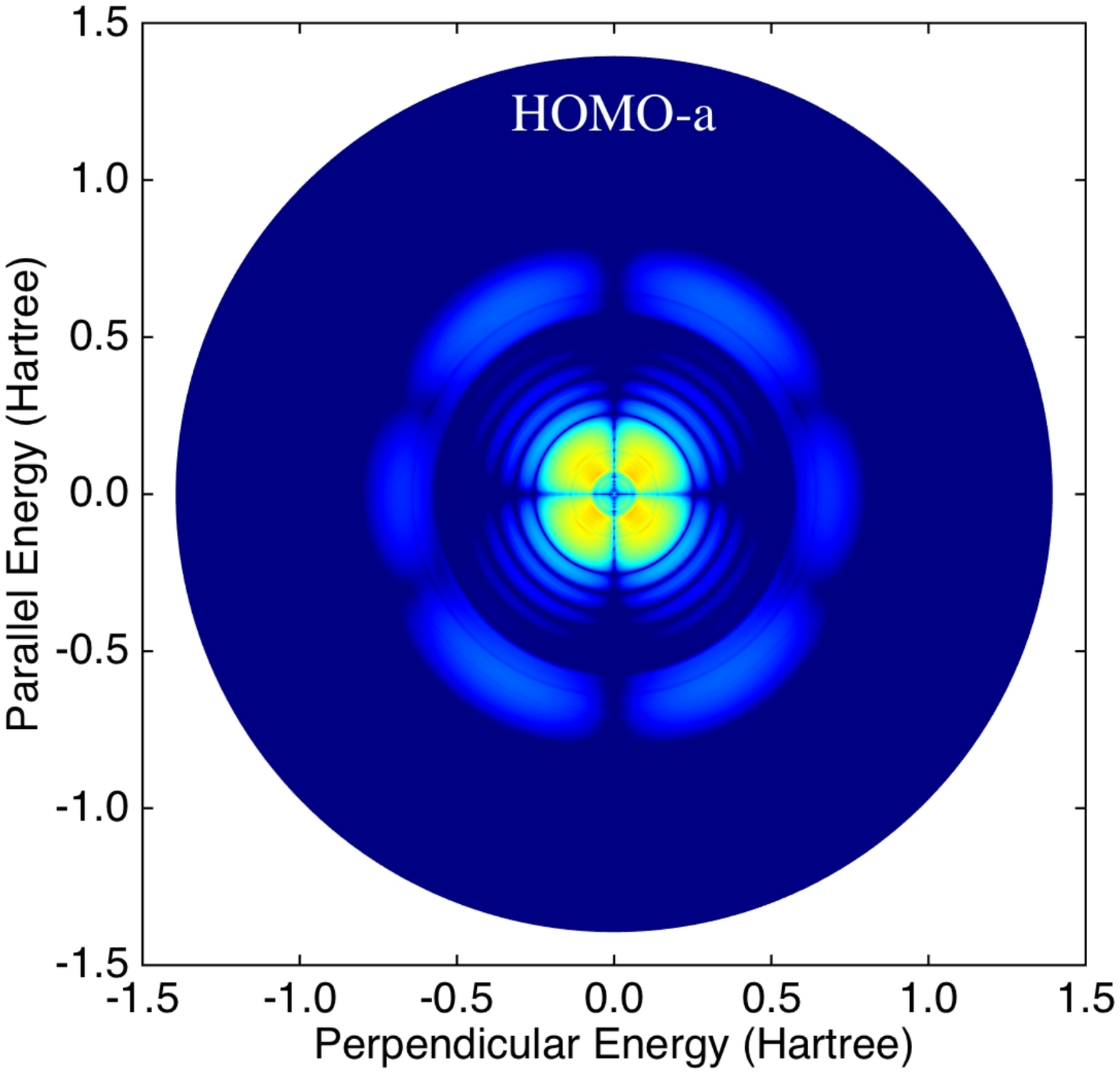}
            \hfill\includegraphics[width=6cm,viewport=12 135 583 707]{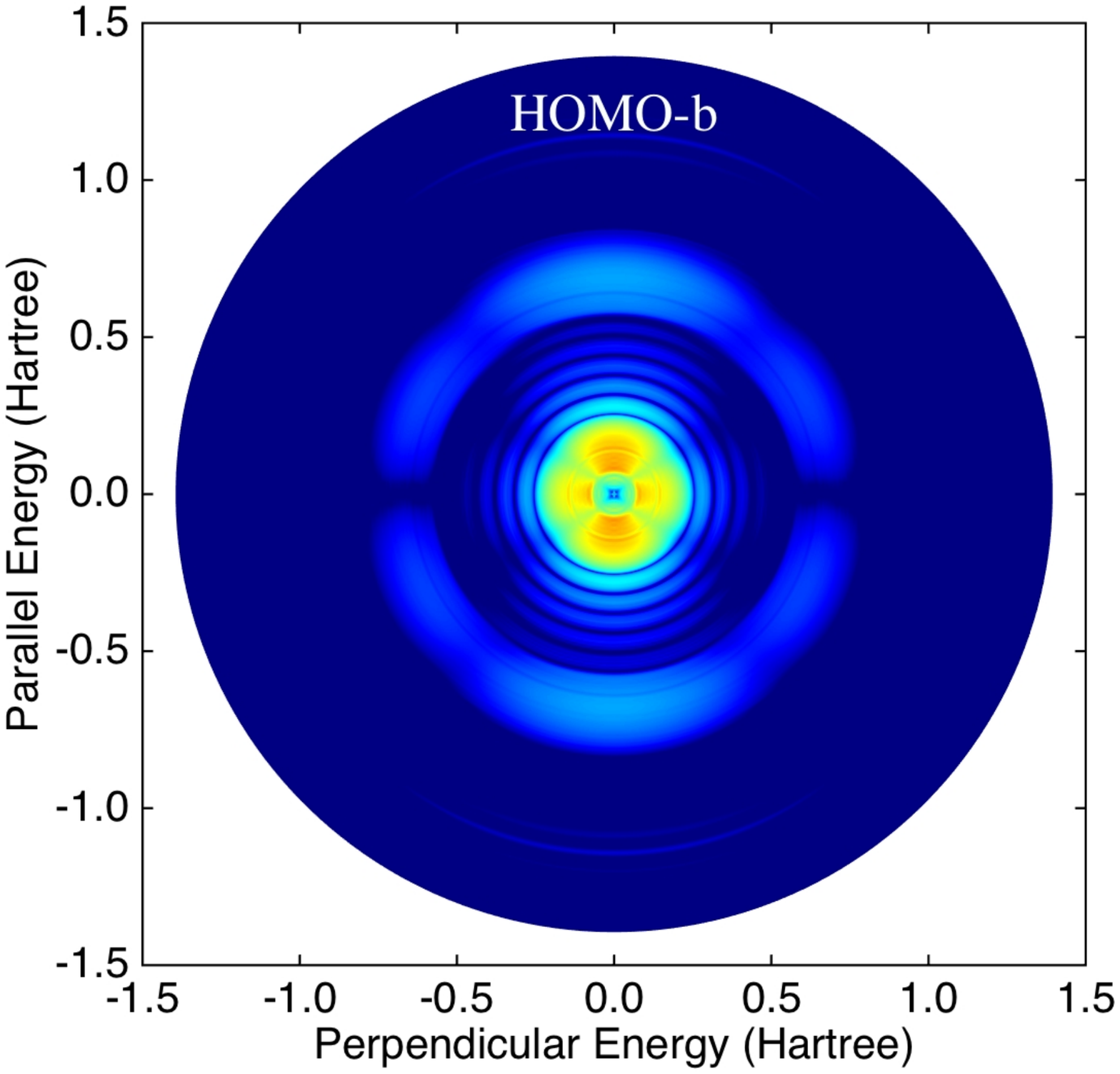}
	    \hfill}
\centerline{\hfill\includegraphics[width=6cm,viewport=12 135 583 707]{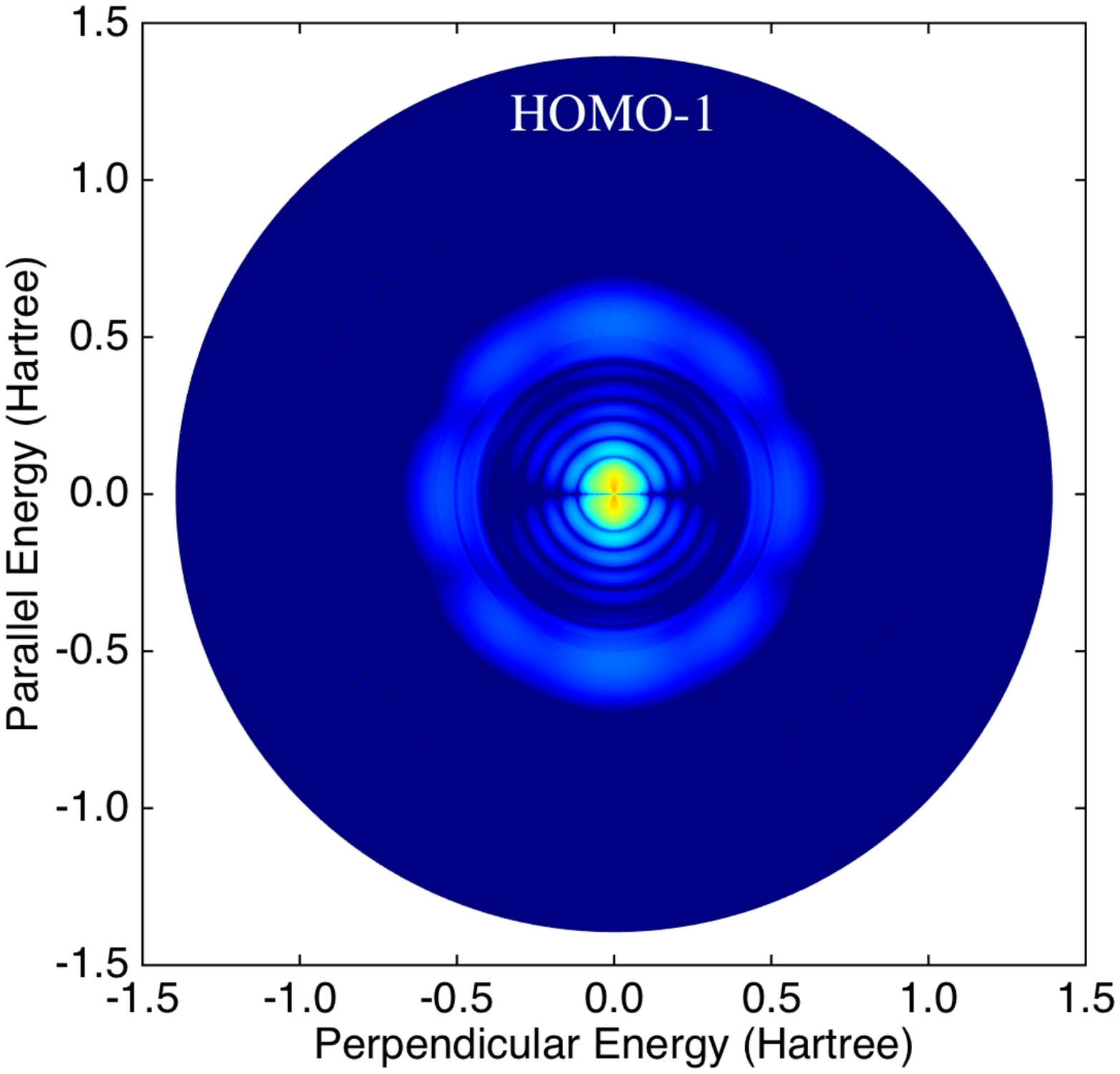}
            \hfill\includegraphics[width=6cm,viewport=12 135 583 707]{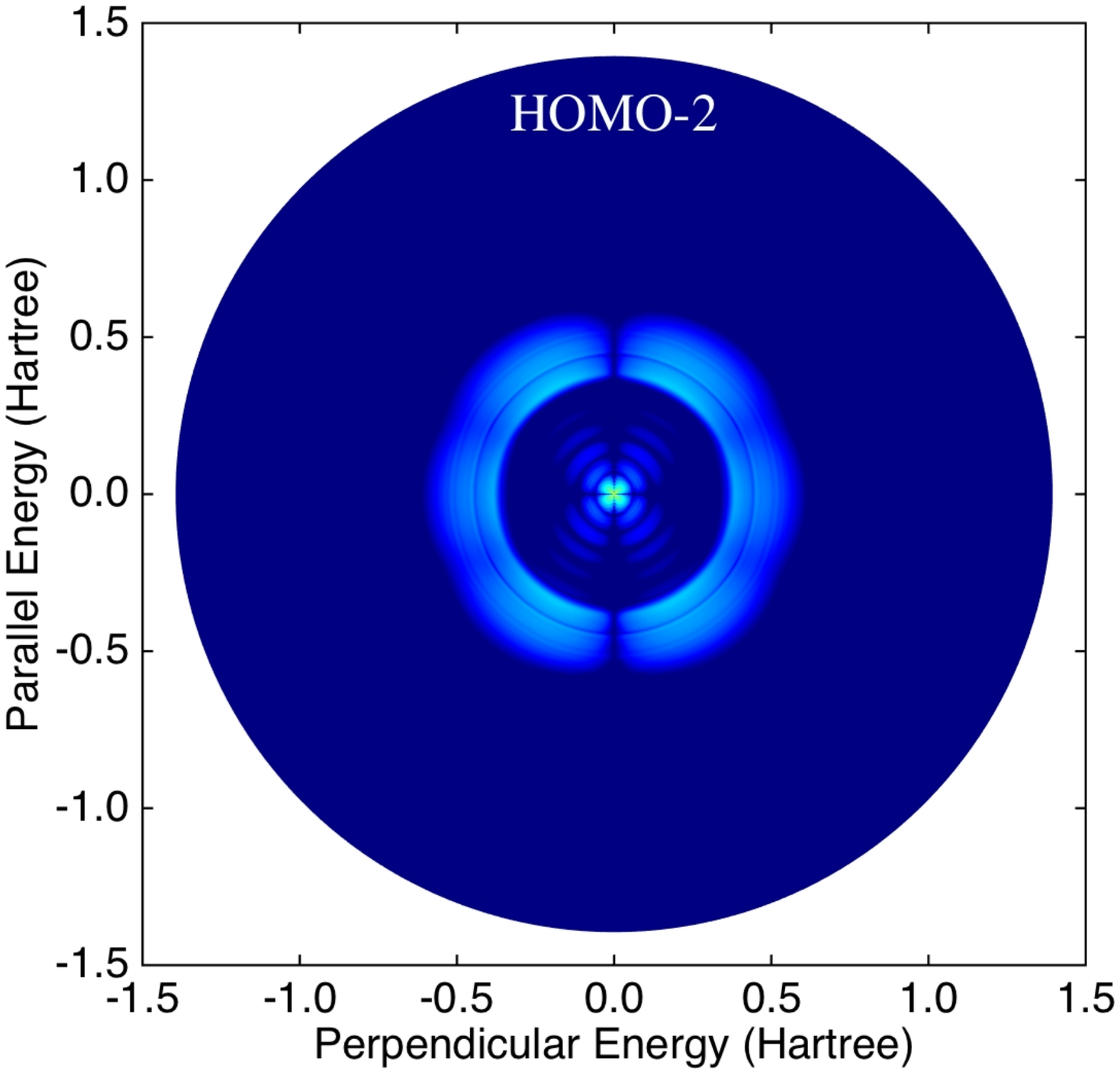}
            \hfill\includegraphics[width=6cm,viewport=12 135 583 707]{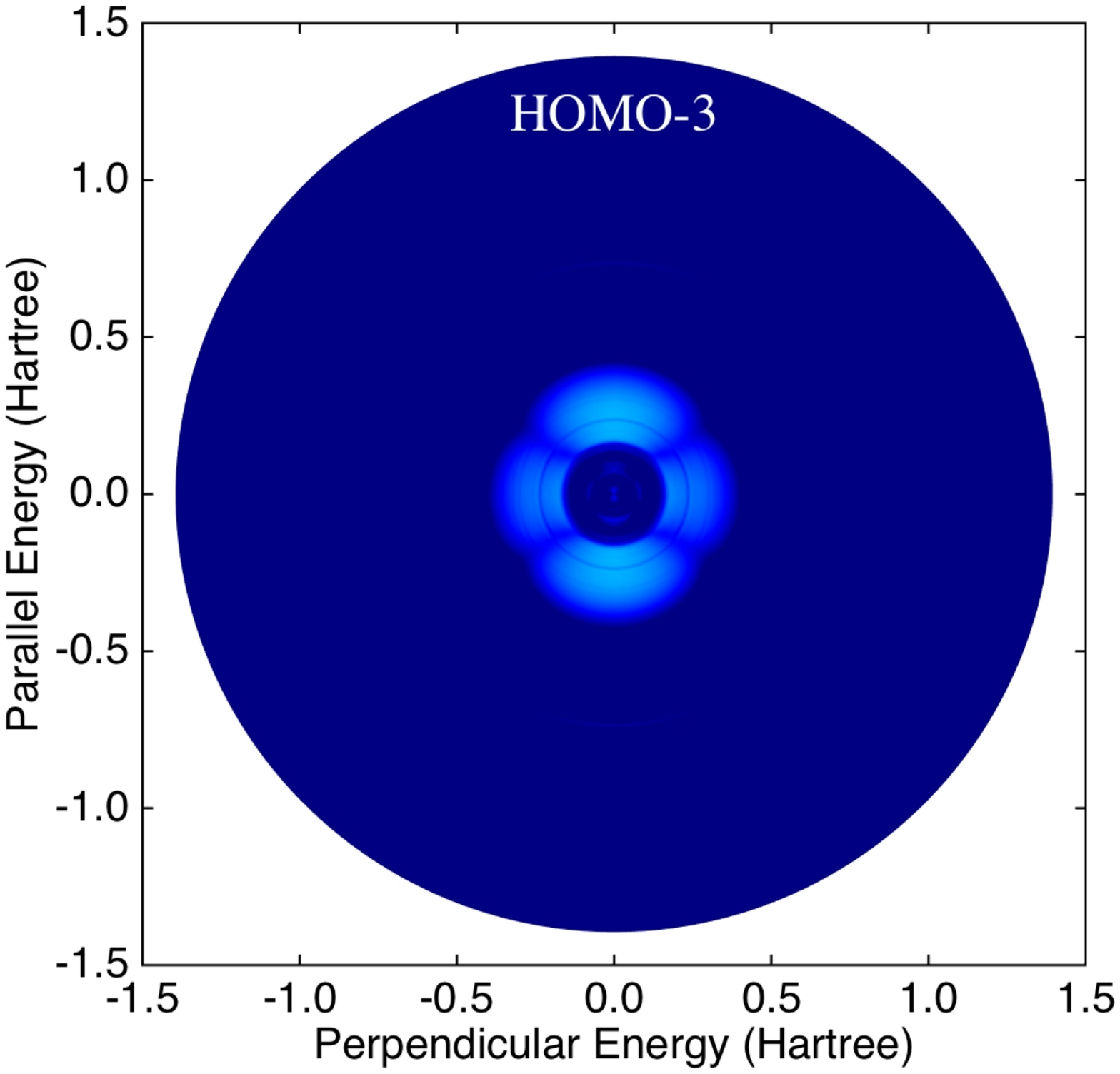}
	    \hfill}
\caption{Angularly-resolved photoelectron spectra for acetylene after interaction with an 8-cycle linearly 
polarized VUV laser pulse having a wavelength of $\lambda=$ 82~nm (photon energy $=$ 0.5557~Ha) 
and a peak intensity of $I = $ \intensity{1.0}{12}. The laser polarization direction is perpendicular to 
the molecular axis (the molecule lies along the $x$-axis). The full photoelectron spectra in 3D has been 
integrated in the azimuthal direction for both the negative $x$-axis and the positive $x$-axis. The left 
semi-circle of each plot gives the results for the negative $x$-axis while the right semi-circle of each plot 
gives the results for the positive $x$-axis. Due to symmetry both quadrants give the same response.}
\label{fig:figure7}
\end{figure*}

\begin{figure*}
\centerline{\hfill\includegraphics[width=8cm,viewport=12 34 527 391]{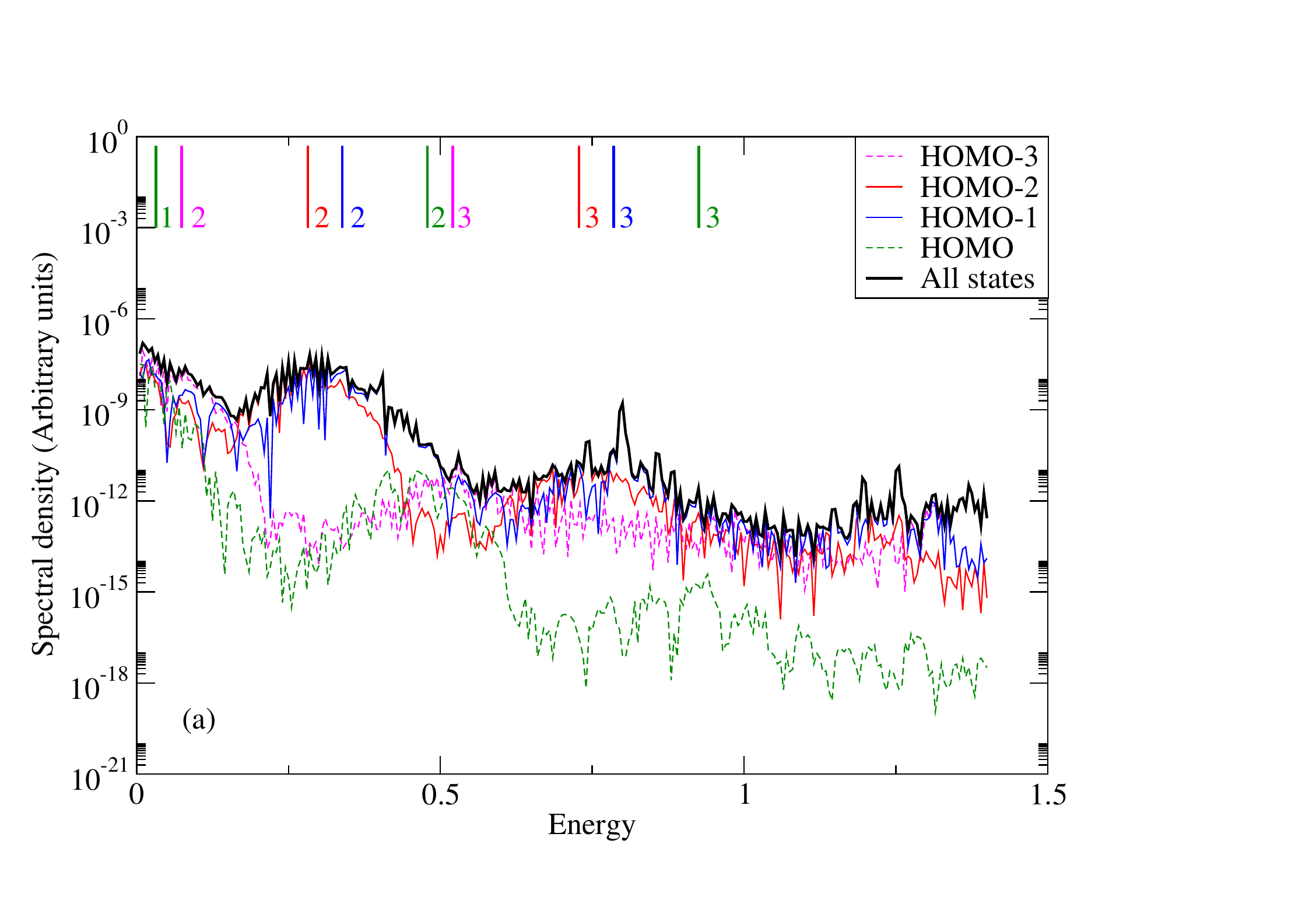}
            \hfill\includegraphics[width=8cm,viewport=12 34 527 391]{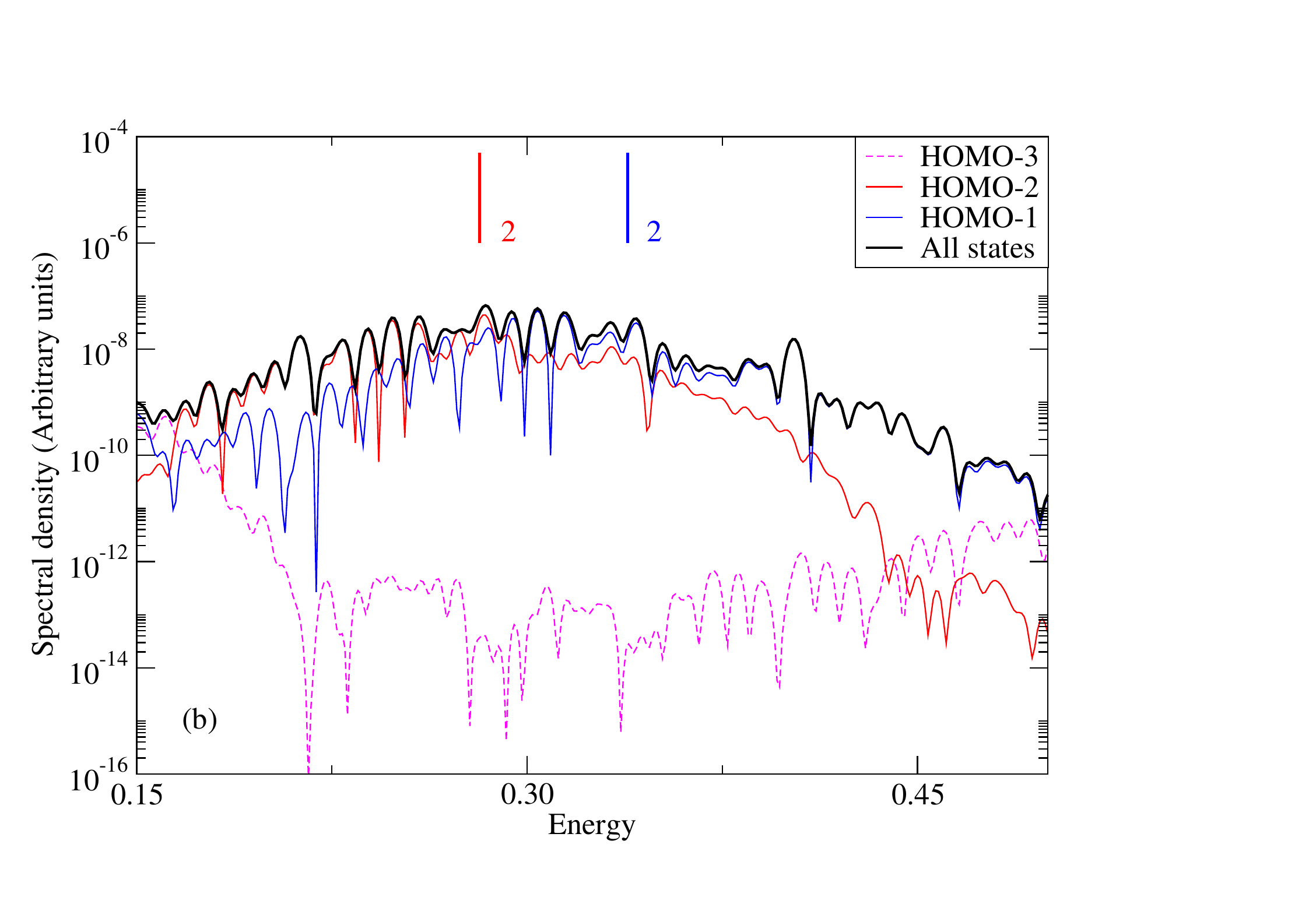}
	    \hfill}
\caption{Photoelectron spectra for acetylene after interaction with an 8-cycle linearly polarized
VUV laser pulse having a wavelength of $\lambda=$ 102~nm (photon energy $=$ 0.4467~Ha) and a peak intensity 
of $I = $ \intensity{1.0}{12}. The laser polarization direction is parallel to the molecular axis. In the 
parallel orientation both HOMO orbitals have the same response and so we only show one. In plot (a) we present the full
spectrum while plot (b) is zoomed in on energies in the range 0.15~Ha to 0.5~Ha. On each plot we also 
show the excess energy associated with vertical ionization due to absorption of one, two or three photons 
from each Kohn-Sham state.}
\label{fig:figure8}
\end{figure*}

\begin{figure*}
\centerline{\hfill\includegraphics[width=6cm,viewport=12 135 583 707]{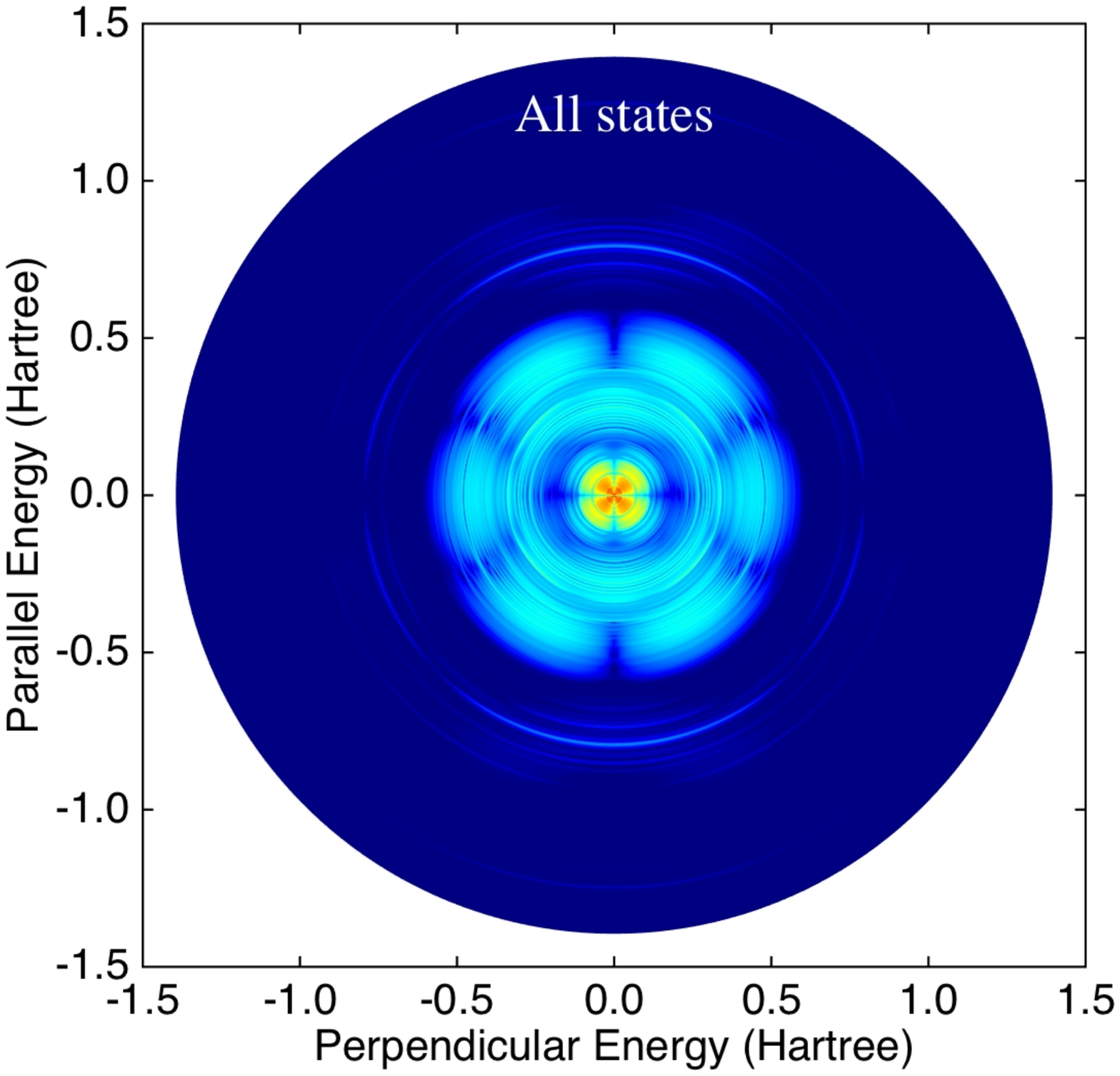}
            \hfill\includegraphics[width=6cm,viewport=12 135 583 707]{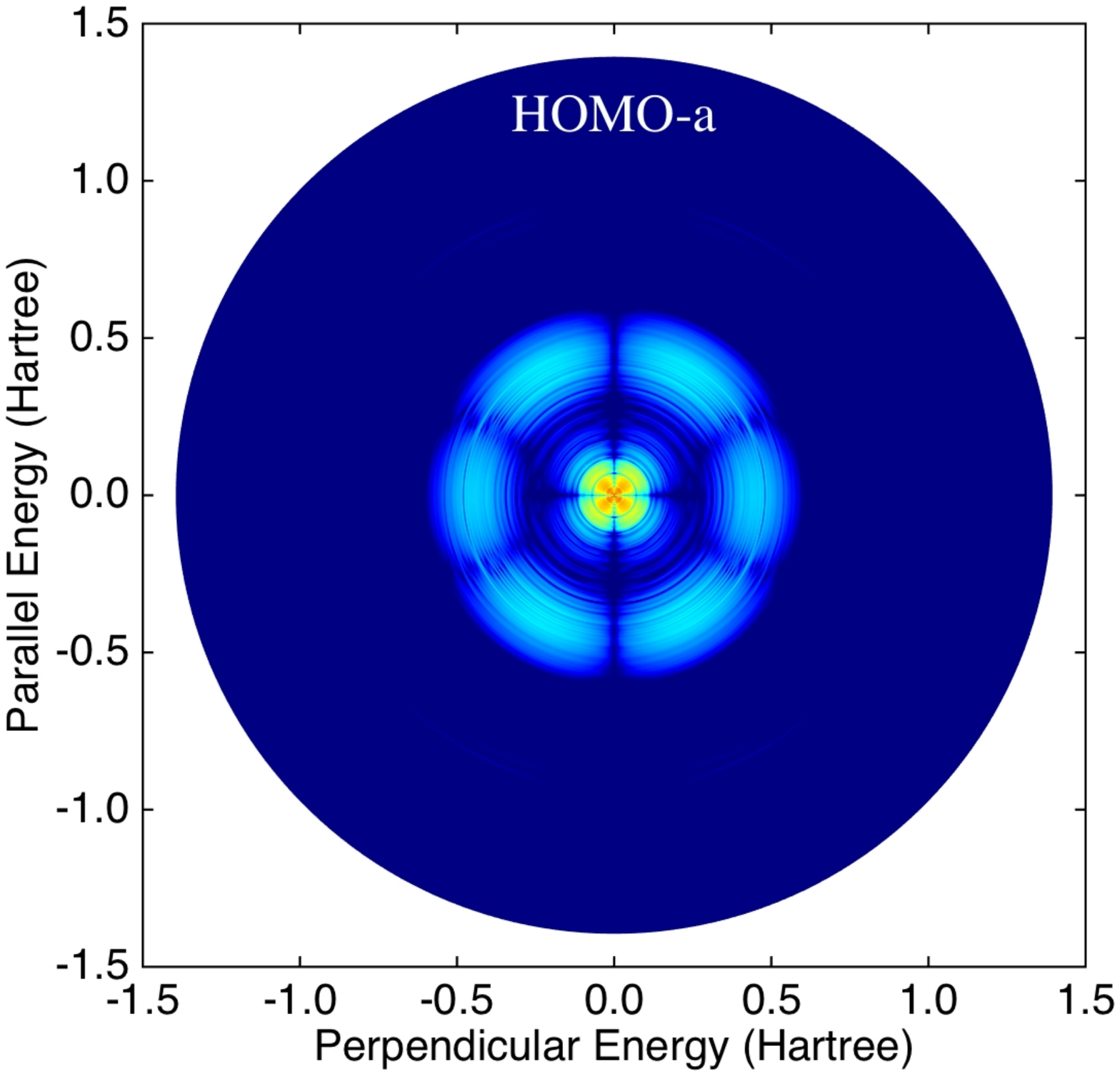}
            \hfill\includegraphics[width=6cm,viewport=12 135 583 707]{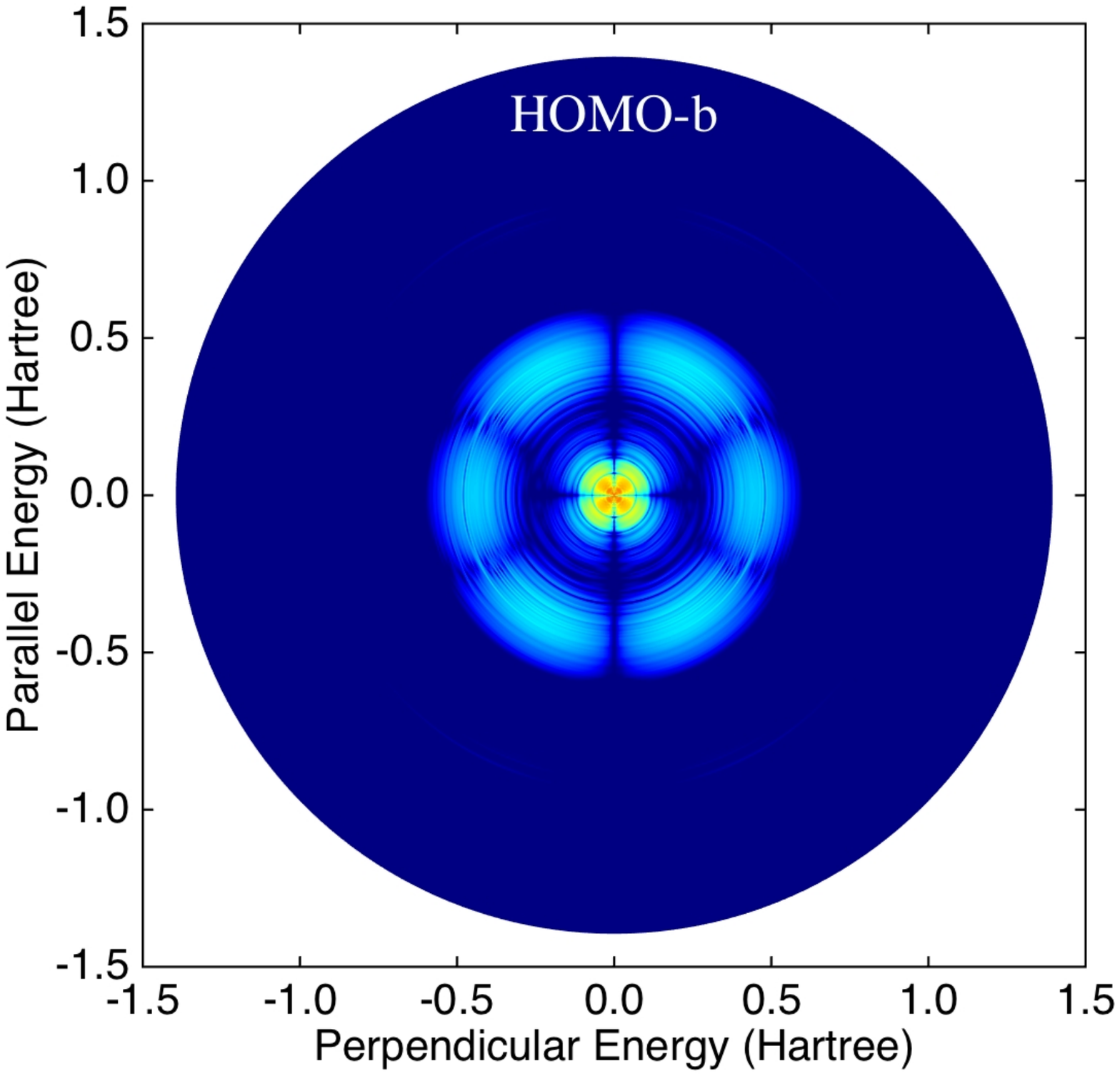}
	    \hfill}
\centerline{\hfill\includegraphics[width=6cm,viewport=12 135 583 707]{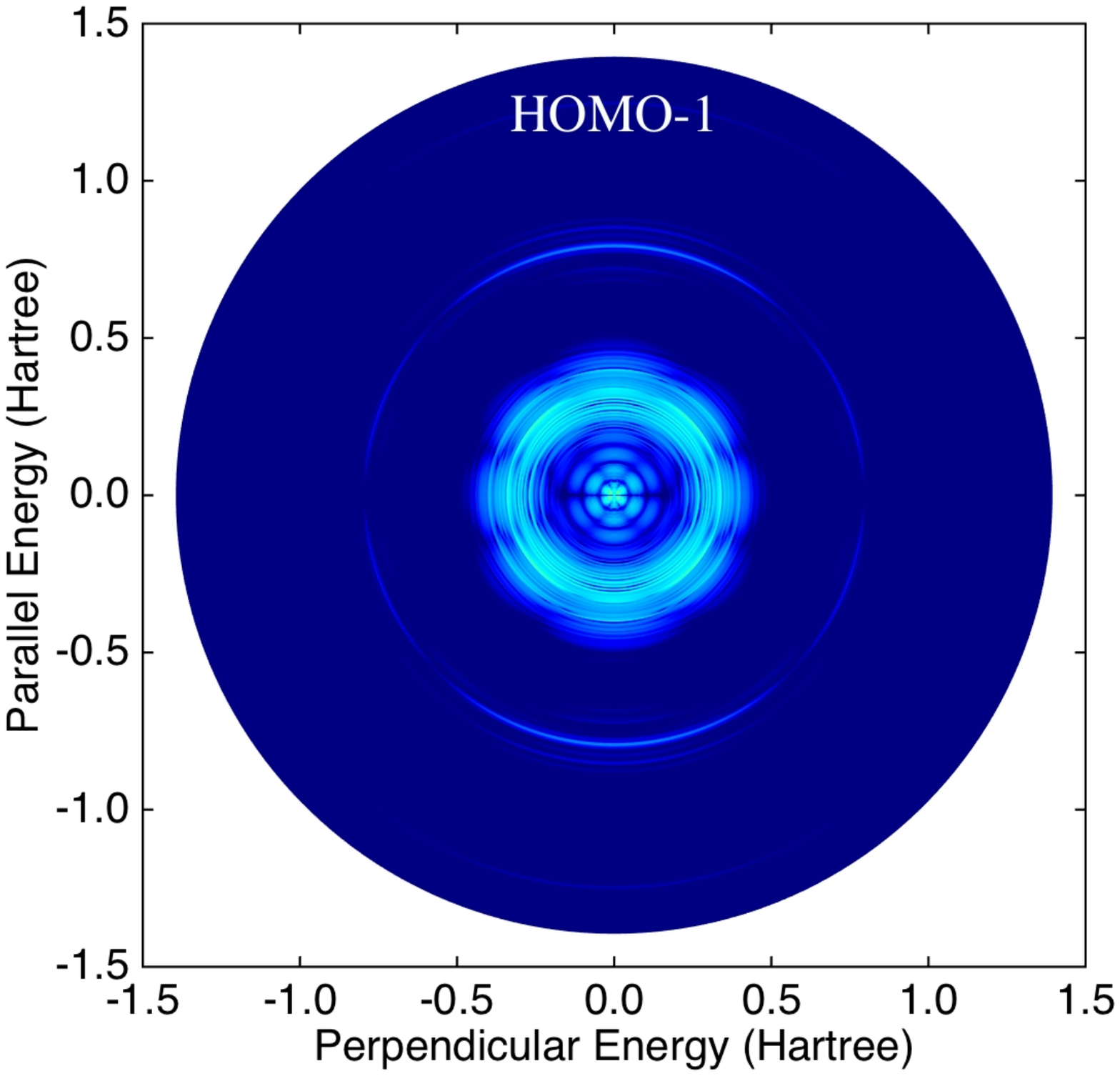}
            \hfill\includegraphics[width=6cm,viewport=12 135 583 707]{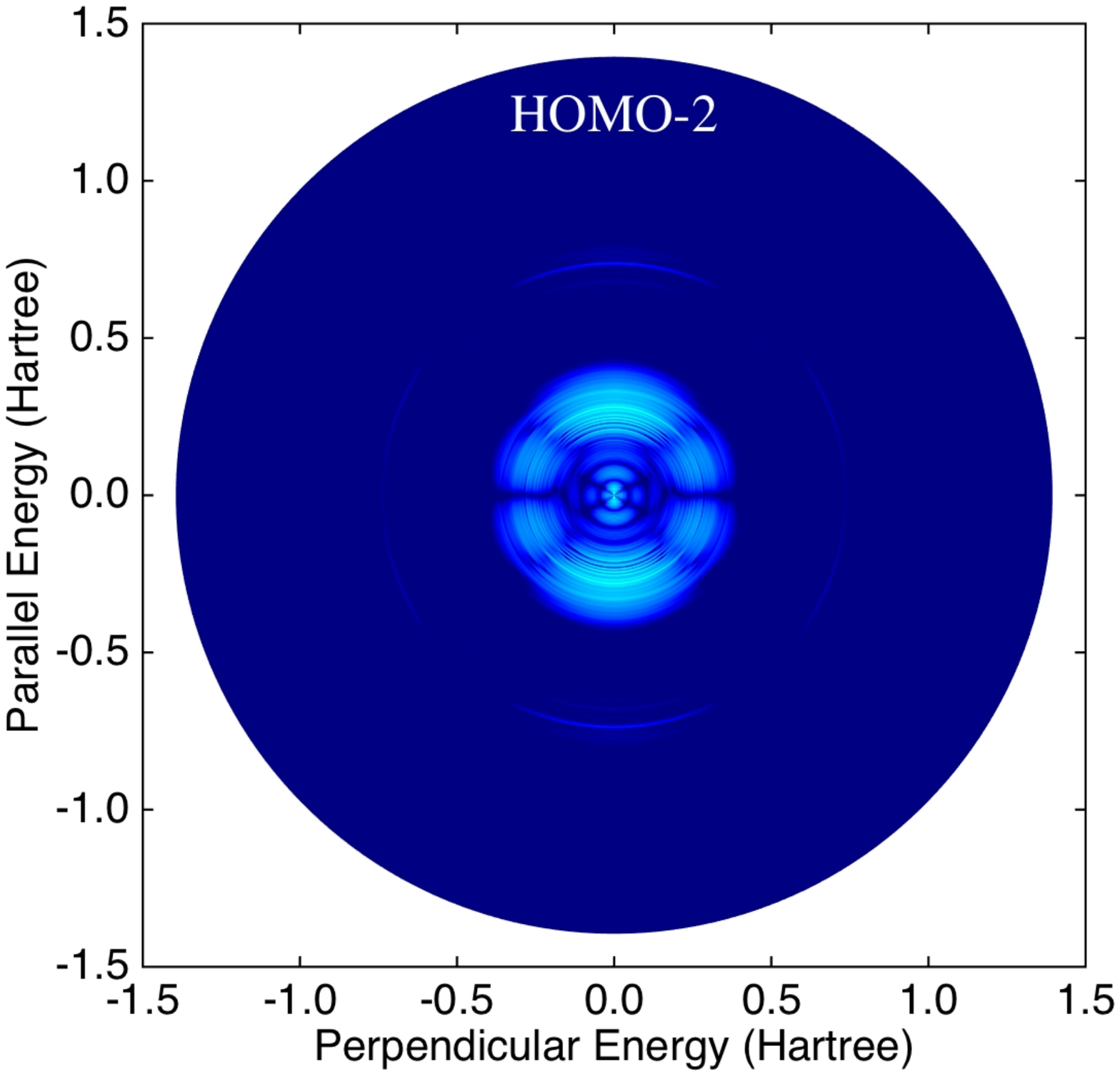}
            \hfill\includegraphics[width=6cm,viewport=12 135 583 707]{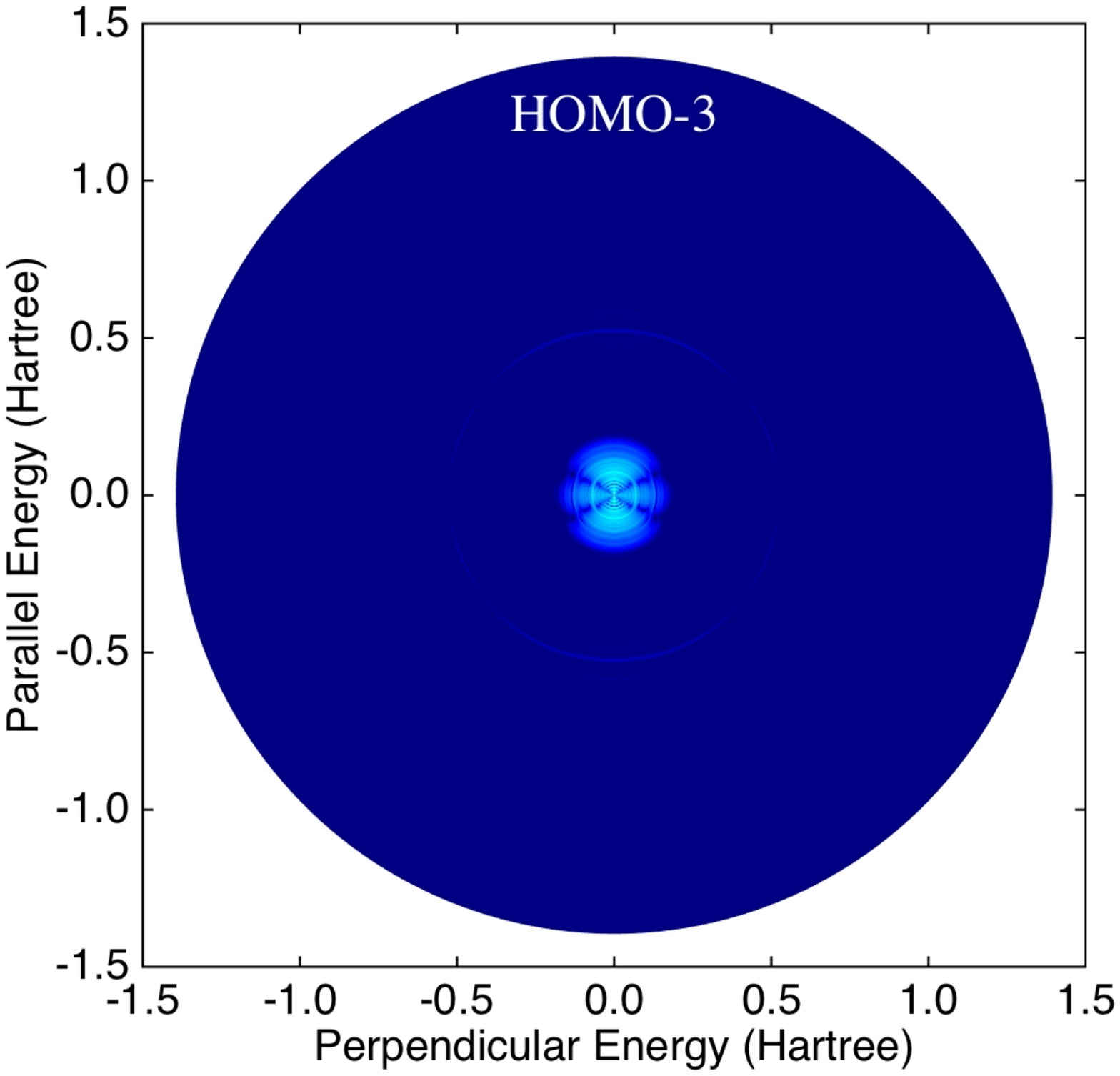}
	    \hfill}
\caption{Angularly-resolved photoelectron spectra for acetylene after interaction with an 8-cycle linearly 
polarized VUV laser pulse having a wavelength of $\lambda=$ 102~nm (photon energy $=$ 0.4467~Ha) 
and a peak intensity of $I = $ \intensity{1.0}{12}. The laser polarization direction is parallel to 
the molecular axis (the molecular axis lies along the $z$-axis). The full photoelectron spectra in 3D has been 
integrated in the azimuthal direction for both the negative $x$-axis and the positive $x$-axis. The left 
semi-circle of each plot gives the results for the negative $x$-axis while the right semi-circle of each plot 
gives the results for the positive $x$-axis. Due to symmetry both quadrants give the same response.}
\label{fig:figure9}
\end{figure*}

\subsection{Photoelectron spectra using VUV laser pulses}
\label{sec:acetylene_pes}
The previous results clearly show how excited states affect HHG in acetylene. Since these excitations were 
initiated by the interaction with a VUV pump pulse we now consider how excited states affect ionization. We 
will calculate photoelectron spectra for acetylene ionized by linearly polarized VUV laser pulses. In 
particular, two wavelengths will be considered: $\lambda =$ 82 nm and $\lambda=$ 102~nm. The shorter wavelength 
provides enough energy to ionize directly from the HOMO-1 orbital with one photon while the longer drives the 
$3\sigma_g\rightarrow 3\sigma_u$ transition with one photon. For these calculations the grid extents were 
$|x| = |y| \leq 76.8$\,a$_0$ and $|z| \leq 90.8$\,a$_0$. The surface was placed at $R_s =$ 60~$a_0$ and the 
maximum angular momentum quantum number used in Eq.~(\ref{eq:tsurff_comm}) was $l_{\max} = 100$. The Kohn-Sham equations
were propagated for 20~fs after the laser pulse finishes.

For the shorter wavelength, we consider the interaction of acetylene with an 8-cycle linearly polarized VUV laser pulse 
having a wavelength of $\lambda=$ 82~nm and a peak intensity of $I = $ \intensity{1.0}{12}. Fig.~\ref{fig:figure5} presents 
photoelectron spectra for two orientations of the molecule with the field. In Fig.~\ref{fig:figure5}(a) the pulse is aligned 
parallel to the molecular axis while in Fig.~\ref{fig:figure5}(b) the pulse is aligned perpendicular to the molecular axis. 
For this laser intensity only a few photons will be absorbed. The results show two broad peaks associated with one- and 
two-photon absorption. The width of each peak is consistent with the the bandwidth of the pulse. As well as plotting the 
full spectra (black lines) we also plot the contribution of each Kohn-Sham orbital. In the parallel orientation the HOMO 
orbitals have negligible response. For the one-photon peak we see that the HOMO-1 dominates whereas for the two-photon peak 
both the HOMO-1 and HOMO-2 contribute: in the lower energy region the HOMO-2 dominates while for the higher energy region the 
HOMO-1 dominates. This arises due to the energy gap between these two orbitals (marked on the graph). In the perpendicular 
case, the response is quite different. Firstly, the overall probability density is an order of magnitude smaller than in the 
parallel orientation. This is to be expected from the results in Fig.~\ref{fig:figure2}. Secondly, one of the HOMO orbitals 
has a significant response while the HOMO-2 orbital has little contribution to the spectrum. Lastly, the HOMO-1 still 
contributes strongly to the spectrum. Due to the large energy gap between the HOMO and HOMO-1 one- and two-photon peaks are 
very broad. Indeed, especially for the one-photon peak, the demarcation between each orbital's contribution can clearly be 
seen.

Fig.~\ref{fig:figure6} presents the angularly-resolved spectrum for the parallel orientation while Fig.~\ref{fig:figure7} 
presents the angularly-resolved spectrum for the perpendicular orientation. In each figure we plot the full spectrum and the 
contribution from each Kohn-Sham orbital. The axes refer to the energy parallel to the pulse polarization direction and that 
perpendicular to it. Consider the parallel alignment results in Fig.~\ref{fig:figure6}. We can clearly see the one- and 
two-photon contribution from each state. For the HOMO orbitals electrons in the one-photon peak are emitted at 45$^{o}$ to the 
laser polarization while they are emitted at 35$^{o}$ and 90$^{o}$ for the two-photon peak. Consider the HOMO-1 and HOMO-2 
orbitals. For the one-photon peak the HOMO-1 electrons are emitted at 55$^{o}$ while for the HOMO-2 electrons are mainly along 
the laser polarization. For the two-photon peak electrons are emitted predominantly along the polarization direction. For the 
HOMO-1 we also see some ionization at 55$^{o}$ and 90$^{o}$ whereas for the HOMO-2 see some emission at 70$^{o}$. Now consider 
the perpendicular alignment results in Fig.~\ref{fig:figure7}. In this case we see that the response of the HOMO orbitals 
dominate. The HOMO-b responds predominantly and we see that electrons are emitted along the pulse polarization direction for 
both one- and two-photon processes. For the one-photon peak, HOMO-1 electrons are preferentially emitted along the pulse 
direction while HOMO-2 electrons are emitted at 55$^{o}$.

For the longer wavelength, consider the interaction of acetylene with an 8-cycle linearly polarized VUV laser pulse having a 
wavelength of $\lambda=$ 102~nm and a peak intensity of $I = $ \intensity{1.0}{12}. Fig.~\ref{fig:figure8} presents 
photoelectron spectra for the parallel orientation. From our HHG results we know that excited states are important for this 
orientation. Fig.~\ref{fig:figure8}(a) presents the full spectrum and the contribution of each Kohn-Sham orbital over the full 
energy range considered. While the basic features of the spectra are similar to those observed at $\lambda =$ 82 nm, there are 
a number of important differences. Firstly, the spectral density is lower. This is to be expected as one-photon absorption results 
in excitation of the HOMO-1 and HOMO-2 orbitals. More importantly, if we consider the two-photon peak we see that is highly 
structured. Fig.~\ref{fig:figure8}(b) zooms in on this peak and presents the spectrum calculated on a much finer energy mesh. 
The structure is a signature of the excited states populated during the absorption of one photon before subsequent ionization 
by a second photon. We clearly see a wide range of Rydberg states populated as well as several resonance features. For the 
HOMO-1 contribution this is generally assigned to the $3\sigma_g\rightarrow np\sigma_u$ Rydberg series whereas for the HOMO-2
it is assigned to the $2\sigma_u\rightarrow ns\sigma_g$ Rydberg series. Fig.~\ref{fig:figure9} presents the angularly-resolved 
spectrum. Again we see that the response is quite different to that observed in Figure~\ref{fig:figure6}. In particular, if we 
consider the contribution of the HOMO-1 orbital we see that the two-photon peak has little angular variation. For the HOMO-2 
orbital, on the other hand, electrons are predominantly emitted along the laser polarization direction.

\section{Conclusions}
\label{sec:conclusions}
We have presented calculations of ionization of acetylene by VUV laser pulses using a TDDFT approach and showed the important role
that excited state dynamics have in the response. In particular we showed that orientation effects have a dramatic effect on the
ionization dynamics. In this case enhanced ionization was found to occur whenever the laser is aligned parallel to the molecular
axis. This is opposite to the situation found whenever high-intensity infra-red (IR) laser pulses are considered. In addition, by 
resonantly tuning the laser wavelength to these excited states, ionization and harmonic generation are significantly modified. 
In a recent paper we showed that HHG from acetylene can be enhanced whenever the molecule is excited to an autoionizing 
state~\cite{mulholland:2017}. Here, we show that this behaviour is due to resonant excitation by considering a range of VUV 
wavelengths. Lastly, we have calculated angularly-resolved PES and showed how the spectrum of emitted electrons changes as the 
wavelength is tuned to various excitations of the molecule.

The excited states probed in this work incorporate resonances and autoionizing states. Such features are common to a wide range 
of molecules and should therefore alter the dynamical response to intense laser pulses. The method used in this work is of 
sufficient generality to be applied to other molecules: this includes more complicated molecules having fewer symmetries. 
As well as being able to study ionization and harmonic generation processes, the implementation of the t-SURFF method provides 
rich information on the energy distribution of the ionized electrons. A particular avenue for future work is studying circular 
dichroism in chiral molecules.

\section{Acknowledgments} 
AW and PM acknowledges financial support through PhD studentships funded by the UK Engineering and Physical Sciences Research 
Council. DD and AdlC acknowledges financial support from the European Union Initial Training Network CORINF and the embedded 
CSE programme of the ARCHER UK National Supercomputing Service (http://www.archer.ac.uk). This work used the ARCHER UK National 
Supercomputing Service and has been supported by COST Action CM1204 (XLIC). 





\begin{mcitethebibliography}{66}
\providecommand*{\natexlab}[1]{#1}
\providecommand*{\mciteSetBstSublistMode}[1]{}
\providecommand*{\mciteSetBstMaxWidthForm}[2]{}
\providecommand*{\mciteBstWouldAddEndPuncttrue}
  {\def\EndOfBibitem{\unskip.}}
\providecommand*{\mciteBstWouldAddEndPunctfalse}
  {\let\EndOfBibitem\relax}
\providecommand*{\mciteSetBstMidEndSepPunct}[3]{}
\providecommand*{\mciteSetBstSublistLabelBeginEnd}[3]{}
\providecommand*{\EndOfBibitem}{}
\mciteSetBstSublistMode{f}
\mciteSetBstMaxWidthForm{subitem}
{(\emph{\alph{mcitesubitemcount}})}
\mciteSetBstSublistLabelBeginEnd{\mcitemaxwidthsubitemform\space}
{\relax}{\relax}

\bibitem[Krausz and Ivanov(2009)]{krauz:2009}
F.~Krausz and M.~Ivanov, \emph{Rev. Mod. Phys.}, 2009, \textbf{81}, 163\relax
\mciteBstWouldAddEndPuncttrue
\mciteSetBstMidEndSepPunct{\mcitedefaultmidpunct}
{\mcitedefaultendpunct}{\mcitedefaultseppunct}\relax
\EndOfBibitem
\bibitem[L\'{e}pine \emph{et~al.}(2014)L\'{e}pine, Ivanov, and
  Vrakking]{lepine:2014}
F.~L\'{e}pine, M.~Y. Ivanov and M.~J.~J. Vrakking, \emph{Nature Photonics},
  2014, \textbf{8}, 195\relax
\mciteBstWouldAddEndPuncttrue
\mciteSetBstMidEndSepPunct{\mcitedefaultmidpunct}
{\mcitedefaultendpunct}{\mcitedefaultseppunct}\relax
\EndOfBibitem
\bibitem[Joachim \emph{et~al.}(2000)Joachim, Gimzewski, and
  Aviram]{joachim:2000}
C.~Joachim, J.~K. Gimzewski and A.~Aviram, \emph{Nature}, 2000, \textbf{408},
  541\relax
\mciteBstWouldAddEndPuncttrue
\mciteSetBstMidEndSepPunct{\mcitedefaultmidpunct}
{\mcitedefaultendpunct}{\mcitedefaultseppunct}\relax
\EndOfBibitem
\bibitem[Willard and van Orden(2003)]{willard:2003}
D.~M. Willard and A.~van Orden, \emph{Nature Materials}, 2003, \textbf{2},
  575\relax
\mciteBstWouldAddEndPuncttrue
\mciteSetBstMidEndSepPunct{\mcitedefaultmidpunct}
{\mcitedefaultendpunct}{\mcitedefaultseppunct}\relax
\EndOfBibitem
\bibitem[Krausz and Stockman(2014)]{krauz:2014}
F.~Krausz and M.~I. Stockman, \emph{Nature Photonics}, 2014, \textbf{8},
  205\relax
\mciteBstWouldAddEndPuncttrue
\mciteSetBstMidEndSepPunct{\mcitedefaultmidpunct}
{\mcitedefaultendpunct}{\mcitedefaultseppunct}\relax
\EndOfBibitem
\bibitem[Calegari \emph{et~al.}(2014)Calegari, Ayuso, Trabattoni, Belshaw,
  Camillis, Anumula, Frassetto, Poletto, Palacios, Decleva, Greenwood,
  Mart\'{i}n, and Nisoli]{calegari:2014}
F.~Calegari, D.~Ayuso, A.~Trabattoni, L.~Belshaw, S.~D. Camillis, S.~Anumula,
  F.~Frassetto, L.~Poletto, A.~Palacios, P.~Decleva, J.~Greenwood,
  F.~Mart\'{i}n and M.~Nisoli, \emph{Science}, 2014, \textbf{346}, 336\relax
\mciteBstWouldAddEndPuncttrue
\mciteSetBstMidEndSepPunct{\mcitedefaultmidpunct}
{\mcitedefaultendpunct}{\mcitedefaultseppunct}\relax
\EndOfBibitem
\bibitem[Trabattoni \emph{et~al.}(2015)Trabattoni, Klinker,
  Gonz\'{a}lez-V\'{a}zquez, Liu, G.~Sansone, Hochlaf, Klei, Vrakking,
  Mart\'{i}n, Nisoli, and Calegari]{trabattoni:2015}
A.~Trabattoni, M.~Klinker, J.~Gonz\'{a}lez-V\'{a}zquez, C.~Liu, R.~L.
  G.~Sansone, M.~Hochlaf, J.~Klei, M.~J.~J. Vrakking, F.~Mart\'{i}n, M.~Nisoli
  and F.~Calegari, \emph{\prx}, 2015, \textbf{5}, 041053\relax
\mciteBstWouldAddEndPuncttrue
\mciteSetBstMidEndSepPunct{\mcitedefaultmidpunct}
{\mcitedefaultendpunct}{\mcitedefaultseppunct}\relax
\EndOfBibitem
\bibitem[Marangos(2016)]{marangos:2016}
J.~P. Marangos, \emph{\jpbo}, 2016, \textbf{49}, 132001\relax
\mciteBstWouldAddEndPuncttrue
\mciteSetBstMidEndSepPunct{\mcitedefaultmidpunct}
{\mcitedefaultendpunct}{\mcitedefaultseppunct}\relax
\EndOfBibitem
\bibitem[Collin and Delwiche(1967)]{collin:1967}
J.~E. Collin and J.~Delwiche, \emph{Can. J. Chem.}, 1967, \textbf{45},
  1883\relax
\mciteBstWouldAddEndPuncttrue
\mciteSetBstMidEndSepPunct{\mcitedefaultmidpunct}
{\mcitedefaultendpunct}{\mcitedefaultseppunct}\relax
\EndOfBibitem
\bibitem[King and Price(2007)]{king:2007}
S.~J. King and S.~D. Price, \emph{\jcp}, 2007, \textbf{127}, 174307\relax
\mciteBstWouldAddEndPuncttrue
\mciteSetBstMidEndSepPunct{\mcitedefaultmidpunct}
{\mcitedefaultendpunct}{\mcitedefaultseppunct}\relax
\EndOfBibitem
\bibitem[Langhoff \emph{et~al.}(1981)Langhoff, McKoy, Unwin, and
  Bradshaw]{langhoff:1981}
P.~W. Langhoff, B.~V. McKoy, R.~Unwin and A.~M. Bradshaw, \emph{Chem. Phys.
  Lett.}, 1981, \textbf{83}, 270\relax
\mciteBstWouldAddEndPuncttrue
\mciteSetBstMidEndSepPunct{\mcitedefaultmidpunct}
{\mcitedefaultendpunct}{\mcitedefaultseppunct}\relax
\EndOfBibitem
\bibitem[Machado \emph{et~al.}(1982)Machado, Leal, Csanak, McKoy, and
  Langhoff]{machado:1982}
L.~E. Machado, E.~P. Leal, G.~Csanak, B.~V. McKoy and P.~W. Langhoff, \emph{J.
  Elec. Spect. Rel. Phenom.}, 1982, \textbf{25}, 1\relax
\mciteBstWouldAddEndPuncttrue
\mciteSetBstMidEndSepPunct{\mcitedefaultmidpunct}
{\mcitedefaultendpunct}{\mcitedefaultseppunct}\relax
\EndOfBibitem
\bibitem[Holland \emph{et~al.}(1983)Holland, West, Parr, Ederer, Stockbauer,
  Buff, and Dehmer]{holland:1983}
D.~M.~P. Holland, J.~B. West, A.~C. Parr, D.~L. Ederer, R.~Stockbauer, R.~D.
  Buff and J.~L. Dehmer, \emph{\jcp}, 1983, \textbf{78}, 124\relax
\mciteBstWouldAddEndPuncttrue
\mciteSetBstMidEndSepPunct{\mcitedefaultmidpunct}
{\mcitedefaultendpunct}{\mcitedefaultseppunct}\relax
\EndOfBibitem
\bibitem[Yasuike and Yabushita(2000)]{yasuike:2000}
T.~Yasuike and S.~Yabushita, \emph{Chem. Phys. Lett.}, 2000, \textbf{316},
  257\relax
\mciteBstWouldAddEndPuncttrue
\mciteSetBstMidEndSepPunct{\mcitedefaultmidpunct}
{\mcitedefaultendpunct}{\mcitedefaultseppunct}\relax
\EndOfBibitem
\bibitem[Fronzoni \emph{et~al.}(2003)Fronzoni, Stener, and
  Decleva]{fronzoni:2003}
G.~Fronzoni, M.~Stener and P.~Decleva, \emph{Chem. Phys.}, 2003, \textbf{298},
  141\relax
\mciteBstWouldAddEndPuncttrue
\mciteSetBstMidEndSepPunct{\mcitedefaultmidpunct}
{\mcitedefaultendpunct}{\mcitedefaultseppunct}\relax
\EndOfBibitem
\bibitem[Argenti \emph{et~al.}(2012)Argenti, Thomas, Liu, Miron, Lischke,
  Pr\"{u}mper, Sakai, Ouchi, P\"{u}ttner, Sekushin, Tanaka, Hoshino, Tanaka,
  Decleva, Ueda, and Mart\'{i}n]{argenti:2012}
L.~Argenti, T.~D. Thomas, E.~P. X.-J. Liu, C.~Miron, T.~Lischke,
  G.~Pr\"{u}mper, K.~Sakai, T.~Ouchi, R.~P\"{u}ttner, V.~Sekushin, T.~Tanaka,
  M.~Hoshino, H.~Tanaka, P.~Decleva, K.~Ueda and F.~Mart\'{i}n, \emph{New J.
  Phys.}, 2012, \textbf{14}, 033012\relax
\mciteBstWouldAddEndPuncttrue
\mciteSetBstMidEndSepPunct{\mcitedefaultmidpunct}
{\mcitedefaultendpunct}{\mcitedefaultseppunct}\relax
\EndOfBibitem
\bibitem[Ji \emph{et~al.}(2015)Ji, Cui, You, Gong, Song, Lin, Pan, Ding, Zeng,
  He, and Wu]{ji:2015}
Q.~Ji, S.~Cui, X.~You, X.~Gong, Q.~Song, K.~Lin, H.~Pan, J.~Ding, H.~Zeng,
  F.~He and J.~Wu, \emph{\pra}, 2015, \textbf{92}, 043401\relax
\mciteBstWouldAddEndPuncttrue
\mciteSetBstMidEndSepPunct{\mcitedefaultmidpunct}
{\mcitedefaultendpunct}{\mcitedefaultseppunct}\relax
\EndOfBibitem
\bibitem[Gaire \emph{et~al.}(2014)Gaire, Lee, Haxton, Pelz, Bocharova, Sturm,
  Gehrken, Honig, Pitzer, Metz, Kim, Sch\"{o}ffler, D\"{o}rner, Gassert,
  Zeller, Voigtsberger, Cao, Zohrabi, Williams, Gatton, Reedy, Nook,
  M\"{u}ller, Landers, Cocke, Ben-Itzhak, Jahnke, Belkacem, and
  Weber]{gaire:2014}
B.~Gaire, S.~Y. Lee, D.~J. Haxton, P.~M. Pelz, I.~Bocharova, F.~P. Sturm,
  N.~Gehrken, M.~Honig, M.~Pitzer, D.~Metz, H.-K. Kim, M.~Sch\"{o}ffler,
  R.~D\"{o}rner, H.~Gassert, S.~Zeller, J.~Voigtsberger, W.~Cao, M.~Zohrabi,
  J.~Williams, A.~Gatton, D.~Reedy, C.~Nook, T.~M\"{u}ller, A.~L. Landers,
  C.~L. Cocke, I.~Ben-Itzhak, T.~Jahnke, A.~Belkacem and T.~Weber, \emph{\pra},
  2014, \textbf{89}, 013403\relax
\mciteBstWouldAddEndPuncttrue
\mciteSetBstMidEndSepPunct{\mcitedefaultmidpunct}
{\mcitedefaultendpunct}{\mcitedefaultseppunct}\relax
\EndOfBibitem
\bibitem[Gong \emph{et~al.}(2014)Gong, Song, Ji, Pan, Ding, Wu, and
  Zeng]{gong:2014}
X.~Gong, Q.~Song, Q.~Ji, H.~Pan, J.~Ding, J.~Wu and H.~Zeng, \emph{\prl}, 2014,
  \textbf{112}, 243001\relax
\mciteBstWouldAddEndPuncttrue
\mciteSetBstMidEndSepPunct{\mcitedefaultmidpunct}
{\mcitedefaultendpunct}{\mcitedefaultseppunct}\relax
\EndOfBibitem
\bibitem[Zammith \emph{et~al.}(2003)Zammith, Blanchet, Girard, Andersson,
  Sorensen, Hjelte, Bj\"{o}rneholm, Gauyacq, Norin, Mauritsson, and
  L'Huillier]{zamith:2003}
S.~Zammith, V.~Blanchet, B.~Girard, J.~Andersson, S.~L. Sorensen, I.~Hjelte,
  O.~Bj\"{o}rneholm, D.~Gauyacq, J.~Norin, J.~Mauritsson and A.~L'Huillier,
  \emph{\jcp}, 2003, \textbf{119}, 3763\relax
\mciteBstWouldAddEndPuncttrue
\mciteSetBstMidEndSepPunct{\mcitedefaultmidpunct}
{\mcitedefaultendpunct}{\mcitedefaultseppunct}\relax
\EndOfBibitem
\bibitem[Vozzi \emph{et~al.}(2010)Vozzi, Torres, Negro, Brugnera, Siegel,
  Altucci, Velotta, Frassetto, Poletto, Villoresi, Silvestri, Stagira, and
  Marangos]{vozzi:2010}
C.~Vozzi, R.~Torres, M.~Negro, L.~Brugnera, T.~Siegel, C.~Altucci, R.~Velotta,
  F.~Frassetto, L.~Poletto, P.~Villoresi, S.~D. Silvestri, S.~Stagira and J.~P.
  Marangos, \emph{App. Phys. Lett.}, 2010, \textbf{97}, 241103\relax
\mciteBstWouldAddEndPuncttrue
\mciteSetBstMidEndSepPunct{\mcitedefaultmidpunct}
{\mcitedefaultendpunct}{\mcitedefaultseppunct}\relax
\EndOfBibitem
\bibitem[Torres \emph{et~al.}(2010)Torres, Siegel, Brugnera, Procino,
  Underwood, Altucci, Velotta, Springate, Froud, Turcu, Ivanov, Smirnova, and
  Marangos]{torres:2010}
R.~Torres, T.~Siegel, L.~Brugnera, I.~Procino, J.~G. Underwood, C.~Altucci,
  R.~Velotta, E.~Springate, C.~Froud, I.~C.~E. Turcu, M.~Y. Ivanov, O.~Smirnova
  and J.~P. Marangos, \emph{Opt. Express}, 2010, \textbf{18}, 3174\relax
\mciteBstWouldAddEndPuncttrue
\mciteSetBstMidEndSepPunct{\mcitedefaultmidpunct}
{\mcitedefaultendpunct}{\mcitedefaultseppunct}\relax
\EndOfBibitem
\bibitem[Vozzi \emph{et~al.}(2012)Vozzi, Negro, and Stagira]{vozzi:2012}
C.~Vozzi, M.~Negro and S.~Stagira, \emph{J. Mod. Opt.}, 2012, \textbf{59},
  1283\relax
\mciteBstWouldAddEndPuncttrue
\mciteSetBstMidEndSepPunct{\mcitedefaultmidpunct}
{\mcitedefaultendpunct}{\mcitedefaultseppunct}\relax
\EndOfBibitem
\bibitem[Negro \emph{et~al.}(2014)Negro, Devetta, Faccial\'{a}, Silvestri,
  Vozzi, and Stagira]{negro:2014}
M.~Negro, M.~Devetta, D.~Faccial\'{a}, S.~D. Silvestri, C.~Vozzi and
  S.~Stagira, \emph{Faraday Discuss.}, 2014, \textbf{171}, 133\relax
\mciteBstWouldAddEndPuncttrue
\mciteSetBstMidEndSepPunct{\mcitedefaultmidpunct}
{\mcitedefaultendpunct}{\mcitedefaultseppunct}\relax
\EndOfBibitem
\bibitem[Ibrahim \emph{et~al.}(2014)Ibrahim, Wales, Beaulieu, Schmidt,
  Thir\'{e}, Fowe, Bisson, Hebeisen, Wanie, Gigu\'{e}re, Kieffer, Spanner,
  Bandrauk, Sanderson, and Schuurman]{ibrahim:2014}
H.~Ibrahim, B.~Wales, S.~Beaulieu, B.~E. Schmidt, N.~Thir\'{e}, E.~P. Fowe,
  E.~Bisson, C.~T. Hebeisen, V.~Wanie, M.~Gigu\'{e}re, J.-C. Kieffer,
  M.~Spanner, A.~D. Bandrauk, J.~Sanderson and M.~S. Schuurman, \emph{Nature
  Comm.}, 2014, \textbf{5}, 4422\relax
\mciteBstWouldAddEndPuncttrue
\mciteSetBstMidEndSepPunct{\mcitedefaultmidpunct}
{\mcitedefaultendpunct}{\mcitedefaultseppunct}\relax
\EndOfBibitem
\bibitem[Lewenstein \emph{et~al.}(1994)Lewenstein, Balcou, Ivanov, L'Huillier,
  and Corkum]{lewenstein:1994}
M.~Lewenstein, P.~Balcou, M.~Y. Ivanov, A.~L'Huillier and P.~B. Corkum,
  \emph{\pra}, 1994, \textbf{49}, 2117\relax
\mciteBstWouldAddEndPuncttrue
\mciteSetBstMidEndSepPunct{\mcitedefaultmidpunct}
{\mcitedefaultendpunct}{\mcitedefaultseppunct}\relax
\EndOfBibitem
\bibitem[Spanner and Patchkovskii(2009)]{spanner:2009}
M.~Spanner and S.~Patchkovskii, \emph{\pra}, 2009, \textbf{80}, 063411\relax
\mciteBstWouldAddEndPuncttrue
\mciteSetBstMidEndSepPunct{\mcitedefaultmidpunct}
{\mcitedefaultendpunct}{\mcitedefaultseppunct}\relax
\EndOfBibitem
\bibitem[Smirnova \emph{et~al.}(2009)Smirnova, Patchkovskii, Dudovich,
  Villeneuve, Corkum, and Ivanov]{smirnova1:2009}
O.~Smirnova, Y.~M.~S. Patchkovskii, N.~Dudovich, D.~Villeneuve, P.~Corkum and
  M.~Y. Ivanov, \emph{Nature}, 2009, \textbf{460}, 972\relax
\mciteBstWouldAddEndPuncttrue
\mciteSetBstMidEndSepPunct{\mcitedefaultmidpunct}
{\mcitedefaultendpunct}{\mcitedefaultseppunct}\relax
\EndOfBibitem
\bibitem[Le \emph{et~al.}(2009)Le, Lucchese, Tonzani, Morishita, and
  Lin]{le:2009}
A.-T. Le, R.~R. Lucchese, S.~Tonzani, T.~Morishita and C.~D. Lin, \emph{\pra},
  2009, \textbf{80}, 013401\relax
\mciteBstWouldAddEndPuncttrue
\mciteSetBstMidEndSepPunct{\mcitedefaultmidpunct}
{\mcitedefaultendpunct}{\mcitedefaultseppunct}\relax
\EndOfBibitem
\bibitem[Lin \emph{et~al.}(2010)Lin, Le, Morishita, and Lucchese]{lin:2010}
C.~D. Lin, A.-T. Le, T.~Morishita and R.~R. Lucchese, \emph{\jpbo}, 2010,
  \textbf{43}, 122001\relax
\mciteBstWouldAddEndPuncttrue
\mciteSetBstMidEndSepPunct{\mcitedefaultmidpunct}
{\mcitedefaultendpunct}{\mcitedefaultseppunct}\relax
\EndOfBibitem
\bibitem[Remacle and Levine(2006)]{remacle:2006}
F.~Remacle and R.~D. Levine, \emph{Proc. Nat. Acad. Sci.}, 2006, \textbf{103},
  6793\relax
\mciteBstWouldAddEndPuncttrue
\mciteSetBstMidEndSepPunct{\mcitedefaultmidpunct}
{\mcitedefaultendpunct}{\mcitedefaultseppunct}\relax
\EndOfBibitem
\bibitem[L\"{u}nnermann \emph{et~al.}(2008)L\"{u}nnermann, Kuleff, and
  Cederbaum]{lunnermann:2008}
S.~L\"{u}nnermann, A.~I. Kuleff and L.~S. Cederbaum, \emph{\jcp}, 2008,
  \textbf{129}, 104305\relax
\mciteBstWouldAddEndPuncttrue
\mciteSetBstMidEndSepPunct{\mcitedefaultmidpunct}
{\mcitedefaultendpunct}{\mcitedefaultseppunct}\relax
\EndOfBibitem
\bibitem[Thachuk \emph{et~al.}(1996)Thachuk, Ivanov, and Wardlaw]{thachuk:1996}
M.~Thachuk, M.~Y. Ivanov and D.~M. Wardlaw, \emph{\jcp}, 1996, \textbf{105},
  4094\relax
\mciteBstWouldAddEndPuncttrue
\mciteSetBstMidEndSepPunct{\mcitedefaultmidpunct}
{\mcitedefaultendpunct}{\mcitedefaultseppunct}\relax
\EndOfBibitem
\bibitem[Mitri\'{c} \emph{et~al.}(2009)Mitri\'{c}, Petersen, and
  Bona\v{c}i\'{c}-Kouteck\'{y}]{mitric:2009}
R.~Mitri\'{c}, J.~Petersen and V.~Bona\v{c}i\'{c}-Kouteck\'{y}, \emph{\pra},
  2009, \textbf{79}, 053416\relax
\mciteBstWouldAddEndPuncttrue
\mciteSetBstMidEndSepPunct{\mcitedefaultmidpunct}
{\mcitedefaultendpunct}{\mcitedefaultseppunct}\relax
\EndOfBibitem
\bibitem[Richter \emph{et~al.}(2011)Richter, Marquetand,
  Gonz\'{a}lez-V\'{a}zquez, Sola, and Gonz\'{a}lez]{richter:2011}
M.~Richter, P.~Marquetand, J.~Gonz\'{a}lez-V\'{a}zquez, I.~Sola and
  L.~Gonz\'{a}lez, \emph{J. Chem. Theory Comput.}, 2011, \textbf{7}, 1253\relax
\mciteBstWouldAddEndPuncttrue
\mciteSetBstMidEndSepPunct{\mcitedefaultmidpunct}
{\mcitedefaultendpunct}{\mcitedefaultseppunct}\relax
\EndOfBibitem
\bibitem[Runge and Gross(1984)]{runge:1984}
E.~Runge and E.~K.~U. Gross, \emph{\prl}, 1984, \textbf{52}, 997\relax
\mciteBstWouldAddEndPuncttrue
\mciteSetBstMidEndSepPunct{\mcitedefaultmidpunct}
{\mcitedefaultendpunct}{\mcitedefaultseppunct}\relax
\EndOfBibitem
\bibitem[Kunert and Schmidt(2003)]{kunert:2003}
T.~Kunert and R.~Schmidt, \emph{Euro. Phys. J. D}, 2003, \textbf{25}, 15\relax
\mciteBstWouldAddEndPuncttrue
\mciteSetBstMidEndSepPunct{\mcitedefaultmidpunct}
{\mcitedefaultendpunct}{\mcitedefaultseppunct}\relax
\EndOfBibitem
\bibitem[Calvayrac \emph{et~al.}(2000)Calvayrac, Reinhard, Suraud, and
  Ullrich]{calvayrac:2000}
F.~Calvayrac, P.-G. Reinhard, E.~Suraud and C.~A. Ullrich, \emph{Phys. Rep.},
  2000, \textbf{337}, 493\relax
\mciteBstWouldAddEndPuncttrue
\mciteSetBstMidEndSepPunct{\mcitedefaultmidpunct}
{\mcitedefaultendpunct}{\mcitedefaultseppunct}\relax
\EndOfBibitem
\bibitem[Marques \emph{et~al.}(2003)Marques, Castro, Bertsch, and
  Rubio]{marques:2003}
M.~A.~L. Marques, A.~Castro, G.~F. Bertsch and A.~Rubio, \emph{Comp. Phys.
  Comm.}, 2003, \textbf{151}, 60\relax
\mciteBstWouldAddEndPuncttrue
\mciteSetBstMidEndSepPunct{\mcitedefaultmidpunct}
{\mcitedefaultendpunct}{\mcitedefaultseppunct}\relax
\EndOfBibitem
\bibitem[Dundas(2012)]{dundas:2012}
D.~Dundas, \emph{\jcp}, 2012, \textbf{136}, 194303\relax
\mciteBstWouldAddEndPuncttrue
\mciteSetBstMidEndSepPunct{\mcitedefaultmidpunct}
{\mcitedefaultendpunct}{\mcitedefaultseppunct}\relax
\EndOfBibitem
\bibitem[Wardlow and Dundas(2016)]{wardlow:2016}
A.~Wardlow and D.~Dundas, \emph{\pra}, 2016, \textbf{93}, 023428\relax
\mciteBstWouldAddEndPuncttrue
\mciteSetBstMidEndSepPunct{\mcitedefaultmidpunct}
{\mcitedefaultendpunct}{\mcitedefaultseppunct}\relax
\EndOfBibitem
\bibitem[Troullier and Martins(1991)]{Troullier:1991}
N.~Troullier and J.~L. Martins, \emph{Phys. Rev. B}, 1991, \textbf{43},
  1993\relax
\mciteBstWouldAddEndPuncttrue
\mciteSetBstMidEndSepPunct{\mcitedefaultmidpunct}
{\mcitedefaultendpunct}{\mcitedefaultseppunct}\relax
\EndOfBibitem
\bibitem[Kleinman and Bylander(1982)]{Kleinman:1982}
L.~Kleinman and D.~M. Bylander, \emph{Phys. Rev. Lett.}, 1982, \textbf{48},
  1425\relax
\mciteBstWouldAddEndPuncttrue
\mciteSetBstMidEndSepPunct{\mcitedefaultmidpunct}
{\mcitedefaultendpunct}{\mcitedefaultseppunct}\relax
\EndOfBibitem
\bibitem[Oliveira and Nogueira(2008)]{Oliveira:2008}
M.~Oliveira and F.~Nogueira, \emph{Comp. Phys. Comm.}, 2008, \textbf{178},
  524\relax
\mciteBstWouldAddEndPuncttrue
\mciteSetBstMidEndSepPunct{\mcitedefaultmidpunct}
{\mcitedefaultendpunct}{\mcitedefaultseppunct}\relax
\EndOfBibitem
\bibitem[Burke \emph{et~al.}(2005)Burke, Werschnik, and Gross]{burke:2005}
K.~Burke, J.~Werschnik and E.~K.~U. Gross, \emph{\jcp}, 2005, \textbf{123},
  062206\relax
\mciteBstWouldAddEndPuncttrue
\mciteSetBstMidEndSepPunct{\mcitedefaultmidpunct}
{\mcitedefaultendpunct}{\mcitedefaultseppunct}\relax
\EndOfBibitem
\bibitem[Perdew and Wang(1992)]{Perdew:1992}
J.~P. Perdew and Y.~Wang, \emph{Phys. Rev. B}, 1992, \textbf{45}, 13244\relax
\mciteBstWouldAddEndPuncttrue
\mciteSetBstMidEndSepPunct{\mcitedefaultmidpunct}
{\mcitedefaultendpunct}{\mcitedefaultseppunct}\relax
\EndOfBibitem
\bibitem[Casida \emph{et~al.}(1998)Casida, Jamorski, Casida, and
  Salahub]{casida:1998}
M.~E. Casida, C.~Jamorski, K.~C. Casida and D.~R. Salahub, \emph{\jcp}, 1998,
  \textbf{108}, 4439\relax
\mciteBstWouldAddEndPuncttrue
\mciteSetBstMidEndSepPunct{\mcitedefaultmidpunct}
{\mcitedefaultendpunct}{\mcitedefaultseppunct}\relax
\EndOfBibitem
\bibitem[Legrand \emph{et~al.}(2002)Legrand, Suraud, and
  Reinhard]{legrand:2002}
C.~Legrand, E.~Suraud and P.-G. Reinhard, \emph{\jpbo}, 2002, \textbf{35},
  1115\relax
\mciteBstWouldAddEndPuncttrue
\mciteSetBstMidEndSepPunct{\mcitedefaultmidpunct}
{\mcitedefaultendpunct}{\mcitedefaultseppunct}\relax
\EndOfBibitem
\bibitem[Krueger and Maitra(2009)]{krueger:2009}
A.~J. Krueger and N.~T. Maitra, \emph{Phys. Chem. Chem. Phys.}, 2009,
  \textbf{11}, 4655\relax
\mciteBstWouldAddEndPuncttrue
\mciteSetBstMidEndSepPunct{\mcitedefaultmidpunct}
{\mcitedefaultendpunct}{\mcitedefaultseppunct}\relax
\EndOfBibitem
\bibitem[Dundas(2004)]{dundas:2004}
D.~Dundas, \emph{J. Phys. B: At. Mol. Opt. Phys.}, 2004, \textbf{37},
  2883\relax
\mciteBstWouldAddEndPuncttrue
\mciteSetBstMidEndSepPunct{\mcitedefaultmidpunct}
{\mcitedefaultendpunct}{\mcitedefaultseppunct}\relax
\EndOfBibitem
\bibitem[Arnoldi(1951)]{Arnoldi:1951}
W.~E. Arnoldi, \emph{Q. Appl. Math}, 1951, \textbf{9}, 17\relax
\mciteBstWouldAddEndPuncttrue
\mciteSetBstMidEndSepPunct{\mcitedefaultmidpunct}
{\mcitedefaultendpunct}{\mcitedefaultseppunct}\relax
\EndOfBibitem
\bibitem[Smyth \emph{et~al.}(1998)Smyth, Parker, and Taylor.]{Smyth:1998}
E.~S. Smyth, J.~S. Parker and K.~Taylor., \emph{Comp. Phys. Comm.}, 1998,
  \textbf{114}, 1\relax
\mciteBstWouldAddEndPuncttrue
\mciteSetBstMidEndSepPunct{\mcitedefaultmidpunct}
{\mcitedefaultendpunct}{\mcitedefaultseppunct}\relax
\EndOfBibitem
\bibitem[Ullrich(2000)]{ullrich:2000}
C.~A. Ullrich, \emph{J. Mol. Struc. - Theochem}, 2000, \textbf{501}, 315\relax
\mciteBstWouldAddEndPuncttrue
\mciteSetBstMidEndSepPunct{\mcitedefaultmidpunct}
{\mcitedefaultendpunct}{\mcitedefaultseppunct}\relax
\EndOfBibitem
\bibitem[Burnett \emph{et~al.}(1992)Burnett, Reed, Cooper, and
  Knight]{burnett:1992}
K.~Burnett, V.~C. Reed, J.~Cooper and P.~L. Knight, \emph{\pra}, 1992,
  \textbf{45}, 3347\relax
\mciteBstWouldAddEndPuncttrue
\mciteSetBstMidEndSepPunct{\mcitedefaultmidpunct}
{\mcitedefaultendpunct}{\mcitedefaultseppunct}\relax
\EndOfBibitem
\bibitem[Chu and Groenenboom(2016)]{chu:2016}
X.~Chu and G.~C. Groenenboom, \emph{\pra}, 2016, \textbf{93}, 013422\relax
\mciteBstWouldAddEndPuncttrue
\mciteSetBstMidEndSepPunct{\mcitedefaultmidpunct}
{\mcitedefaultendpunct}{\mcitedefaultseppunct}\relax
\EndOfBibitem
\bibitem[Tao and Scrinzi(2012)]{Tao:2012}
L.~Tao and A.~Scrinzi, \emph{New J. Phys.}, 2012, \textbf{12}, 013021\relax
\mciteBstWouldAddEndPuncttrue
\mciteSetBstMidEndSepPunct{\mcitedefaultmidpunct}
{\mcitedefaultendpunct}{\mcitedefaultseppunct}\relax
\EndOfBibitem
\bibitem[Scrinzi(2012)]{Scrinzi:2012}
A.~Scrinzi, \emph{New J. Phys.}, 2012, \textbf{14}, 085008\relax
\mciteBstWouldAddEndPuncttrue
\mciteSetBstMidEndSepPunct{\mcitedefaultmidpunct}
{\mcitedefaultendpunct}{\mcitedefaultseppunct}\relax
\EndOfBibitem
\bibitem[Yue and Madsen(2014)]{Yue:2014}
L.~Yue and L.~B. Madsen, \emph{Phys. Rev. A}, 2014, \textbf{90}, 063408\relax
\mciteBstWouldAddEndPuncttrue
\mciteSetBstMidEndSepPunct{\mcitedefaultmidpunct}
{\mcitedefaultendpunct}{\mcitedefaultseppunct}\relax
\EndOfBibitem
\bibitem[Yue and Madsen(2013)]{Yue:2013}
L.~Yue and L.~B. Madsen, \emph{Phys. Rev. A}, 2013, \textbf{88}, 063420\relax
\mciteBstWouldAddEndPuncttrue
\mciteSetBstMidEndSepPunct{\mcitedefaultmidpunct}
{\mcitedefaultendpunct}{\mcitedefaultseppunct}\relax
\EndOfBibitem
\bibitem[Karamatskou \emph{et~al.}(2014)Karamatskou, Pabst, Chen, and
  Santra]{Karamatskou:2014}
A.~Karamatskou, S.~Pabst, Y.-J. Chen and R.~Santra, \emph{Phys. Rev. A}, 2014,
  \textbf{89}, 033415\relax
\mciteBstWouldAddEndPuncttrue
\mciteSetBstMidEndSepPunct{\mcitedefaultmidpunct}
{\mcitedefaultendpunct}{\mcitedefaultseppunct}\relax
\EndOfBibitem
\bibitem[Wopperer \emph{et~al.}(2017)Wopperer, Giovannini, and
  Rubio]{wopperer:2016}
P.~Wopperer, U.~D. Giovannini and A.~Rubio, \emph{Eur. Phys. J. B, 2017 90, 5},
  2017, \textbf{90}, 5\relax
\mciteBstWouldAddEndPuncttrue
\mciteSetBstMidEndSepPunct{\mcitedefaultmidpunct}
{\mcitedefaultendpunct}{\mcitedefaultseppunct}\relax
\EndOfBibitem
\bibitem[Zielinski(2016)]{zielinski:2016}
A.~Zielinski, PhD Thesis, Ludwig Maximilians Universit\"{a}t, 2016\relax
\mciteBstWouldAddEndPuncttrue
\mciteSetBstMidEndSepPunct{\mcitedefaultmidpunct}
{\mcitedefaultendpunct}{\mcitedefaultseppunct}\relax
\EndOfBibitem
\bibitem[Lekien and Marsden(2005)]{lekien:2005}
F.~Lekien and J.~Marsden, \emph{Int. J. Num. Meth. Eng.}, 2005, \textbf{63},
  455\relax
\mciteBstWouldAddEndPuncttrue
\mciteSetBstMidEndSepPunct{\mcitedefaultmidpunct}
{\mcitedefaultendpunct}{\mcitedefaultseppunct}\relax
\EndOfBibitem
\bibitem[{POpSiCLE library}()]{popsicle}
{POpSiCLE library}, https://ccpforge.cse.rl.ac.uk/gf/project/popsicle\relax
\mciteBstWouldAddEndPuncttrue
\mciteSetBstMidEndSepPunct{\mcitedefaultmidpunct}
{\mcitedefaultendpunct}{\mcitedefaultseppunct}\relax
\EndOfBibitem
\bibitem[Wells and Lucchese(1999)]{wells:1999}
M.~C. Wells and P.~R. Lucchese, \emph{\jcp}, 1999, \textbf{111}, 6290\relax
\mciteBstWouldAddEndPuncttrue
\mciteSetBstMidEndSepPunct{\mcitedefaultmidpunct}
{\mcitedefaultendpunct}{\mcitedefaultseppunct}\relax
\EndOfBibitem
\bibitem[Mulholland and Dundas()]{mulholland:2017}
P.~Mulholland and D.~Dundas, \prl, submitted\relax
\mciteBstWouldAddEndPuncttrue
\mciteSetBstMidEndSepPunct{\mcitedefaultmidpunct}
{\mcitedefaultendpunct}{\mcitedefaultseppunct}\relax
\EndOfBibitem
\end{mcitethebibliography}

\providecommand*{\mcitethebibliography}{\thebibliography}
\csname @ifundefined\endcsname{endmcitethebibliography}
{\let\endmcitethebibliography\endthebibliography}{}

\end{document}